\documentclass[reprint, superscriptaddress, nofootinbib, altaffilsymbol, aps, prc, floatfix, longbibliography]{revtex4-1}
\usepackage[colorlinks=true,pdfstartview=FitV,linkcolor=blue,citecolor=blue,urlcolor=blue,bookmarks=false]{hyperref}
\usepackage[usenames,dvipsnames]{color}
\usepackage[sc]{mathpazo}
\usepackage[below]{placeins}
\usepackage{afterpage}
\usepackage[separate-uncertainty,retain-explicit-plus,per-mode = symbol, detect-weight=true, range-phrase=-, range-units=single]{siunitx}
\usepackage{mathtools}
\usepackage{cuted}
\usepackage{multirow}
\usepackage{rotating}
\usepackage[mathlines]{lineno}
\usepackage{graphicx}
\usepackage{paralist}
\usepackage{wrapfig}
\usepackage{appendix}
\usepackage{vruler}
\usepackage{etoolbox}
\usepackage{tikz}
\usepackage{listings}
\usepackage{amsmath}
\usepackage{relsize}
\usepackage{etoolbox}
\usepackage{gensymb}
\usepackage{colortbl}
\usepackage{comment}
\usepackage{nicefrac}
\usepackage[caption=false]{subfig}  

\modulolinenumbers[5]
\newcolumntype{d}[1]{D{.}{\cdot}{#1}}
\newcommand{\isotope}[2]{$^{#2}${#1}}

\DeclareSIUnit\bqkg{Bq/kg}
\DeclareSIUnit\ppt{pg/g}
\DeclareSIUnit\ppb{ng/g}
\DeclareSIUnit\ppm{\ensuremath{\micro}g/g}
\DeclareSIUnit\gpg{g/g}
\DeclareSIUnit\c{$c$}
\DeclareSIUnit\day{day}
\DeclareSIUnit\week{w}
\DeclareSIUnit\year{yr}
\DeclareSIUnit\standard{std}
\DeclareSIUnit\str{sr}

\newcommand{\liquidn}{\text{LN}}
\newcommand{\gasn}{\text{GN}}
\newcommand{\bipo}{\isotope{BiPo}{212}}
\newcommand{\PMT}{\text{PMT}}
\newcommand{\QE}{\text{QE}}
\newcommand{\PNNL}{\text{PNNL}}
\newcommand{\PTFE}{\text{PTFE}}
\newcommand{\PE}{\text{PE}}
\newcommand{\SPE}{\text{SPE}}
\newcommand{\airfly}{\text{AIRFLY}}
\newcommand{\DAQ}{\text{DAQ}}
\newcommand{\STP}{\text{STP}}
\newcommand{\NLTS}{\text{NLTS}}
\newcommand{\LCE}{\text{LCE}}
\newcommand{\RT}{\text{RT}}
\newcommand{\CE}{\text{CE}}
\newcommand{\DE}{\text{DE}}
\newcommand{\precooldowntime}{\SI{0.9}{\day}}
\newcommand{\avegasnitrogenpressure}{\SI{1317 \pm 5}{Torr}}
\newcommand{\cooldowntime}{\SI{2.5}{\day}} 
\newcommand{\aveliquidnitrogenpressure}{\SI{1082 \pm 9}{Torr}}
\newcommand{\aveliquidnitrogentemperature}{\SI{80.5 \pm 0.2}{\kelvin}}
\newcommand{\fillingrate}{\SI{1.7}{SLPM}}
\newcommand{\totalfill}{\SI{2.2}{\liter} of \liquidn}
\newcommand{\totdays}{\SI{10.1}{\day}}
\newcommand{\totdaysvacuum}{\SI{3.1}{\day}}
\newcommand{\avepressuregnvacuum}{\SI{1319}{Torr}}
\newcommand{\avepressurevacuum}{\SI{7e-3}{Torr}}

\newcommand{\daysbottomofcup}{\SI{1.0}{\day}}
\newcommand{\dayspmtlevel}{\SI{1.4}{\day}}

\newcommand{\pmtdiam}{\SI{3}{in}} 
\newcommand{\ptfecupdiam}{\SI{2.5}{in}~ID} 
\newcommand{\ptfecupheight}{\SI{2.5}{in}} 
\newcommand{\pmtviewvolume}{\SI{202}{\centi\metre\cubed}}
\newcommand{\ptfecupbottomwallthickness}{\SI{0.5}{in}} 

\newcommand{\thoronsourceactivity}{\SI{10.2 \pm 0.1}{\kilo\becquerel} of \isotope{Rn}{220} (26 May 2023)}

\newcommand{\Pbtauexpected}{\SI{15.3}{\hour}}
\newcommand{\Pbhalflifeexpected}{\SI{10.6}{\hour}}
\newcommand{\Pbtaumeasured}{\qty[uncertainty-descriptors={stat,sys}]{17.1(0.4)(0.3)}{\hour}}
\newcommand{\BiPotauexpected}{\SI{431.4}{\nano\second}}
\newcommand{\BiPohalflifeexpected}{\SI{299}{\nano\second}}
\newcommand{\BiPotaumeasured}{\qty[uncertainty-descriptors={stat,sys}]{451.1(9.0)(4.5)}{\nano\second}}

\newcommand{\LNrelativeyield}{$Y_{\liquidn}/Y^{\STP}_{\gasn} =$ \num[uncertainty-descriptors={stat,sys}]{0.0142(0.0005)(0.0030)}}
\newcommand{\LNrelativeyieldPercentage}{\SI{1.4}{\%}}
\newcommand{\LNrelativeyieldRT}{$Y_{\liquidn}/Y^{\STP}_{\gasn} = \num[uncertainty-descriptors={stat,sys}]{0.0136(0.0004)(0.0028)}$}

\newcommand{\LNabsoluteyield}{$Y_{\liquidn} = \qty[uncertainty-descriptors={stat,sys}]{2.39(0.08)(0.55)}{PH\per\MeV}$}
\newcommand{\LNabsoluteyieldTOT}{$Y_{\liquidn} = \qty[]{2.39(0.56)}{photons~per~MeV}$}

\newcommand{\muonrate}{\SI{217}{muons/\metre\squared/\second}} 
\newcommand{\spesystematicln}{\SI{2.3}{\%}} %
\newcommand{\spesystematicgn}{\SI{1.5}{\%}} %
\newcommand{\spesystematicgnRT}{\SI{1.8}{\%}} %

\newcommand{\dtexcludedregionPb}{\SI{0.5}{\micro\second}} %

\newcommand{\BiPotaumeasuredgn}{\qty[uncertainty-descriptors={stat,sys}]{423.5(10.5)(1.3:7.8)}{\nano\second}}

\newcommand{\avetemperaturegn}{\SI{240 \pm 6}{\kelvin}} %
\newcommand{\avepressuregn}{\SI{1387\pm29}{Torr}} %
\newcommand{\avetemperaturegnRT}{\SI{297.40 \pm 0.01}{\kelvin}} %
\newcommand{\avepressuregnRT}{\SI{1316 \pm 20}{Torr}} %

\newcommand{\qonemugn}{\qty[uncertainty-descriptors={stat,sys}]{166.9(0.5)(3.8)}{\PE}}
\newcommand{\qonemugnRT}{\qty[uncertainty-descriptors={stat,sys}]{208.8(0.6)(4.9)}{\PE}}
\newcommand{\qonemuln}{\qty[uncertainty-descriptors={stat,sys}]{5.43(0.17)(0.79)}{\PE}}

\newcommand{\Pbtaumeasuredvacuum}{\qty[uncertainty-descriptors={stat,sys}]{15.9(0.4)(0.2)}{\hour}}
\newcommand{\BiPotaumeasuredvacuum}{\qty[uncertainty-descriptors={stat,sys}]{406.4(9.8)(0.8:14.9)}{\nano\second}}

\newcommand{\bkgsystematicvacuum}{\SI{1.9}{\%}} %
\newcommand{\bkgsystematicflat}{\SI{5.4}{\%}} %
\newcommand{\bkgsystematicpemin}{\SI{13}{\%}} %

\newcommand{\airflycorrection}{$\Theta(\SI{1387}{Torr},\SI{240}{\kelvin}) = \qty[]{0.49 \pm 0.02}{}$} %
\newcommand{\airflycorrectionRT}{$\Theta(\SI{1315}{Torr},\SI{297}{\kelvin}) = \qty[]{0.61 \pm 0.02}{}$} %
\newcommand{\pprimeairfly}{$p^{'} = \qty[]{77.8 \pm 2.0}{Torr}$} %
\newcommand{\alphaairfly}{$\alpha = \qty[]{-0.36 \pm 0.08}{}$} %
\newcommand{\GNabsoluteyieldave}{$Y_{\gasn}^{\STP} = \qty[]{169 \pm 15}{PH\per\MeV}$}

\newcommand{\ThetaTpcorrectionbandvariation}{\SI{1.8}{\%}}

\newcommand{\QEliquid}{\SI{45.7 \pm 8.0}{\%}}
\newcommand{\QEgas}{\SI{40.9 \pm 7.1}{\%}}
\newcommand{\QEgasRT}{\SI{39.2 \pm 4.1}{\%}}
\newcommand{\QEratio}{\SI{0.89 \pm 0.13}{}}
\newcommand{\QEratioRT}{\SI{0.86 \pm 0.12}{}}
\newcommand{\QEgrowth}{\SI{-0.030 \pm 0.004}{\%\per\kelvin}}

\newcommand{\LCEgas}{\SI{78 \pm 16}{\%}}
\newcommand{\LCEgasRT}{\SI{82 \pm 12}{\%}}

\newcommand{\fitrangesystematicgn}{\SI{0.5}{\%}} %
\newcommand{\fitrangesystematicgnRT}{\SI{0.6}{\%}} %

\newcommand{\fitstratsystematicgn}{\SI{1.6}{\%}} %
\newcommand{\fitstratsystematicgnRT}{\SI{1.4}{\%}} %
\newcommand{\mcfitgn}{\SI{169.6}{\PE}} %
\newcommand{\mcfitgnRT}{\SI{211.7}{\PE}} %

\begin{document}

\title{Scintillation of liquid nitrogen}
\newcommand{\pnnl}{Pacific Northwest National Laboratory, Richland, WA 99352, USA}
\author{L. Pagani}\email[Corresponding author: ]{luca.pagani@pnnl.gov}\affiliation{\pnnl}
\author{R. Saldanha}\email[Corresponding author: ]{richard.saldanha@pnnl.gov}\affiliation{\pnnl}
\author{B.M. Loer}\affiliation{\pnnl}
\author{G.S. Ortega}\affiliation{\pnnl}
\author{R.A. Bunker}\altaffiliation{Now at SNOLAB, Sudbury, ON, Canada.}\affiliation{\pnnl}
\author{B.T. Foust}\altaffiliation{Now at SRNL, Aiken, SC, 29808.}\affiliation{\pnnl}


\begin{abstract}
Liquid nitrogen is commonly used in cryogenic applications and is a promising medium for the direct immersion cooling of sensors used for nuclear and particle physics experiments. 
The scintillation properties of gaseous nitrogen are well-documented, but little is known about the scintillation of liquid nitrogen. If present, scintillation light from interactions of ambient radioactivity could produce backgrounds for rare event searches such as the direct detection of dark matter. Using a coincidence-tagged alpha decay, we demonstrate that liquid nitrogen exhibits measurable, albeit very faint, scintillation. Assuming the same scintillation wavelength spectrum as gaseous nitrogen, we estimate a relative scintillation yield of \LNrelativeyield\ with respect to gaseous nitrogen at standard temperature and pressure. Considering the average scintillation yield from alpha decays in gaseous nitrogen, this implies a scintillation yield for alpha decays in liquid nitrogen of \LNabsoluteyieldTOT. To our knowledge this is the first measurement of scintillation in liquid nitrogen.
\end{abstract}
\maketitle

\section{Introduction}

Liquid nitrogen (\liquidn) is used as a cryogenic cooling fluid for a wide range of applications. Due to its low boiling point and high atmospheric abundance, it is the coolant of choice for many instruments and devices operating between \SIrange{77}{200}{\kelvin}. Because of its electrically insulating nature and lack of long-lived radioactive isotopes, \liquidn\ is being considered as the coolant for the direct immersion cooling of charge coupled devices (CCDs) within the Oscura dark matter experiment~\cite{aguilararevalo2022oscura, cervantes-vergara2023skipper}. Previous and current generations of CCD dark matter experiments use ultra-high purity copper to thermally couple the CCD sensors to a cryocooler~\cite{damicsearch2016, senseifirst2025, damic2024damicm}. While copper can be produced with extremely low-levels of intrinsic radioactive contamination, cosmogenic activation and plate out of radon progeny on the surface can make significant contributions to the background levels of the experiment. To ensure uniform cooling of the much larger number (\SI{\sim 24000}{}) of CCD sensors while avoiding the use of large amounts of high radiopurity copper, Oscura intends to directly immerse the CCDs in \liquidn, pressurized to roughly \SI{450}{psi} to obtain an operating temperature of \SIrange{120}{140}{\kelvin}. 

One potential concern with direct immersion cooling is the generation of light within the cooling medium itself. CCDs are sensitive to a broad range of wavelengths (\SIrange{200}{1100}{\nano\meter}), and even small amounts of stray light can lead to backgrounds that affect the sensitivity of the dark matter search due to the stringent background requirements of the Oscura experiment (\SI{<0.025}{events\per\keV\per\kg\per\day})~\cite{cervantes-vergara2023skipper}. It is well established that gaseous nitrogen (\gasn) scintillates (see Ref.~\cite{BIRKS1964570} for a historical review) and this scintillation has been used to detect interactions of particles in gas scintillation counters~\cite{conde1967argon, tornow1976properties, saito2023nitrogen} or ultra-high energy cosmic rays in the atmosphere~\cite{fly1985utah, airfly2006airfly, auger2010fluorescence}. However, somewhat surprisingly given its ubiquitous use, there is no published literature on whether liquid nitrogen scintillates. To our knowledge, the only documented mention of scintillation in \liquidn\ occurs in an unpublished internal document from CERN that proposed to use liquid nitrogen as a shield for a solar neutrino detector ~\cite{CERNpreprint} where they were unable to detect any scintillation signal. 
Since no upper limit on the scintillation yield was provided, the objective of this study is to investigate whether \liquidn\ scintillates, and to quantify its yield (either as a value or upper limit) in comparison to \gasn.

\section{Experimental Plan}

Initial measurements with a \liquidn\ cell and surrounding scintillator panels showed significant amounts of light produced by through-going muons (compared to an evacuated cell). However, since the average muon energy on the surface (\SI{\sim 4}{\GeV}) is above the threshold for Cherenkov light production in \liquidn\ (see Tab.~\ref{tab:ethparticles}), it was unclear if the light was produced through scintillation or Cherenkov emission. Electron interactions are not ideal for scintillation measurements as the Cherenkov threshold for electrons is only \SI{\sim 400}{\keV}, and, given the likely small scintillation yield, scintillation from low energy electrons would be hard to detect.
In fact, a Compton-edge feature was not apparent above background in the spectrum of coincidence-tagged \SI{511}{\keV} annihilation gamma interactions. Alphas from radioactive decays have energies that are typically in the \SIrange{4}{9}{\MeV} range, high enough to potentially produce detectable amounts of scintillation light but well below the Cherenkov threshold (see Tab.~\ref{tab:ethparticles}), therefore serving as an ideal probe for investigating scintillation in \liquidn.

Commercially available alpha sources are not certified for use in cryogenic liquids and concerns about radioactive contamination of the experimental setup prevented us from using custom encapsulated sources. We instead used a gaseous \isotope{Rn}{220} source to measure the progeny \isotope{Po}{212} alpha emission in \liquidn\ (see the decay chain shown in Fig.~\ref{fig:rn220decay}). Using a \isotope{Rn}{220} source has several advantages.
\begin{itemize}
    \item \isotope{Po}{212} emits an alpha with \SI{8785}{\keV} of kinetic energy, towards the high end of alpha energies from radioactive decay, making it easier to see a potential scintillation signal.
    \item \isotope{Po}{212} has a short half-life (\SI{299}{\nano\second}) which allows it to be tagged in coincidence with its parent isotope \isotope{Bi}{212}. \isotope{Bi}{212} is a $\beta$ emitter with a Q-value of \SI{2252}{\keV}, above the Cherenkov threshold, and thus is expected to produce a visible signal even in the absence of scintillation.
    \item The parent \isotope{Pb}{212} is sufficiently long-lived (\SI{10.1}{\hour} half-life) to allow it to move through the condensation loop and enter the detector before decaying. 
    \item The entire decay chain of \isotope{Rn}{220} is relatively short-lived such that there is no concern about residual radioactive contamination a few days after the source is isolated from the system. 
\end{itemize}
We note that while the alpha emitted from the decay of \isotope{Po}{212} is not energetic enough to produce Cherenkov radiation directly or through delta electrons (maximum energy of \SI{4.8}{\keV}), it can produce gamma rays through inelastic scattering or nuclear reactions (e.g. \isotope{N}{14}($\alpha$,p)\isotope{O}{17} or \isotope{N}{14}($\alpha, \gamma$)\isotope{F}{18}), bremsstrahlung emission~\cite{boie2007brem}, or electrons through ejection of core shell electrons~\cite{fischbeck1975spectroscopy}. However, as discussed later, these processes are all extremely rare and have very low probabilities of producing Cherenkov radiation.

\begin{table}
    \centering
    \begin{tabular}{ccccc}
        \hline
        Medium & Refractive index & \multicolumn{3}{c}{Cherenkov Threshold [MeV]}\\
        & & Electrons & Muons & Alphas \\
        \hline
        \gasn\ & 1.0003~\cite{Griesmann1999, Borzsonyi2008} & 20.4 & 4220 & $149 \times 10^3$\\
        \liquidn\ & 1.21~\cite{JOHNS1937, Liveing01101893} & 0.40 & 83 & 2938\\
        \PTFE\ & 1.36~\cite{French2009} & 0.24 & 50 & 1772\\
        Fused silica & 1.48~\cite{Tan2005} & 0.18 & 38 & 1329\\
        \hline
    \end{tabular}
    \caption{Energy threshold for the production of Cherenkov light by different particles in \gasn\ and \liquidn, as well as detector materials used in the experimental setup (polytetrafluoroethylene (\PTFE) and fused silica). The indexes of refraction are given at \SI{337}{\nano\metre}, except for \liquidn\ which was averaged between \SI{436}{} to \SI{694}{\nano\metre}.}
    \label{tab:ethparticles}
\end{table}

\section{Experimental Setup}

\subsection{Nitrogen Detector}

The active region of the detector consists of a \ptfecupdiam, \ptfecupheight\ tall \PTFE\ cup, coupled to a \pmtdiam\ photomultiplier tube (\PMT) at the top face. 
The \pmtviewvolume\ active region  was designed to have high light collection efficiency (\LCE), with $>$ \ptfecupbottomwallthickness-thick \PTFE\ walls providing \SI{>92}{\%} diffuse reflectivity over the wavelengths of interest~\cite{Weidner81, Ghosh2020} and the inner diameter restricted to the active photocathode area of the \PMT. 
We used a Hamamatsu R11065 \PMT~\cite{HR11065} with a quantum efficiency (\QE) of \SI{39.2}{\%} at $\lambda_0 = \qty{337}{\nano\meter}$ (see Appx.~\ref{appendix:pmt}), and a voltage divider customized for high linearity. 
The active volume directly receives the condensed cryogen through a filling line positioned at the bottom of the cup.
This allows for the direct injection of dissolved radioactive sources into the sensitive volume of the detector, aiding the measurement of short-lived isotopes. The cup and the \PMT\ are housed in a \SI{5.4}{in}~ID, \SI{24}{in} tall double-walled cryostat with an \SI{8}{in} flanged top that contained all the fluid and electrical feedthroughs.

\begin{figure}[tb]
\centering
\includegraphics[width=\columnwidth]{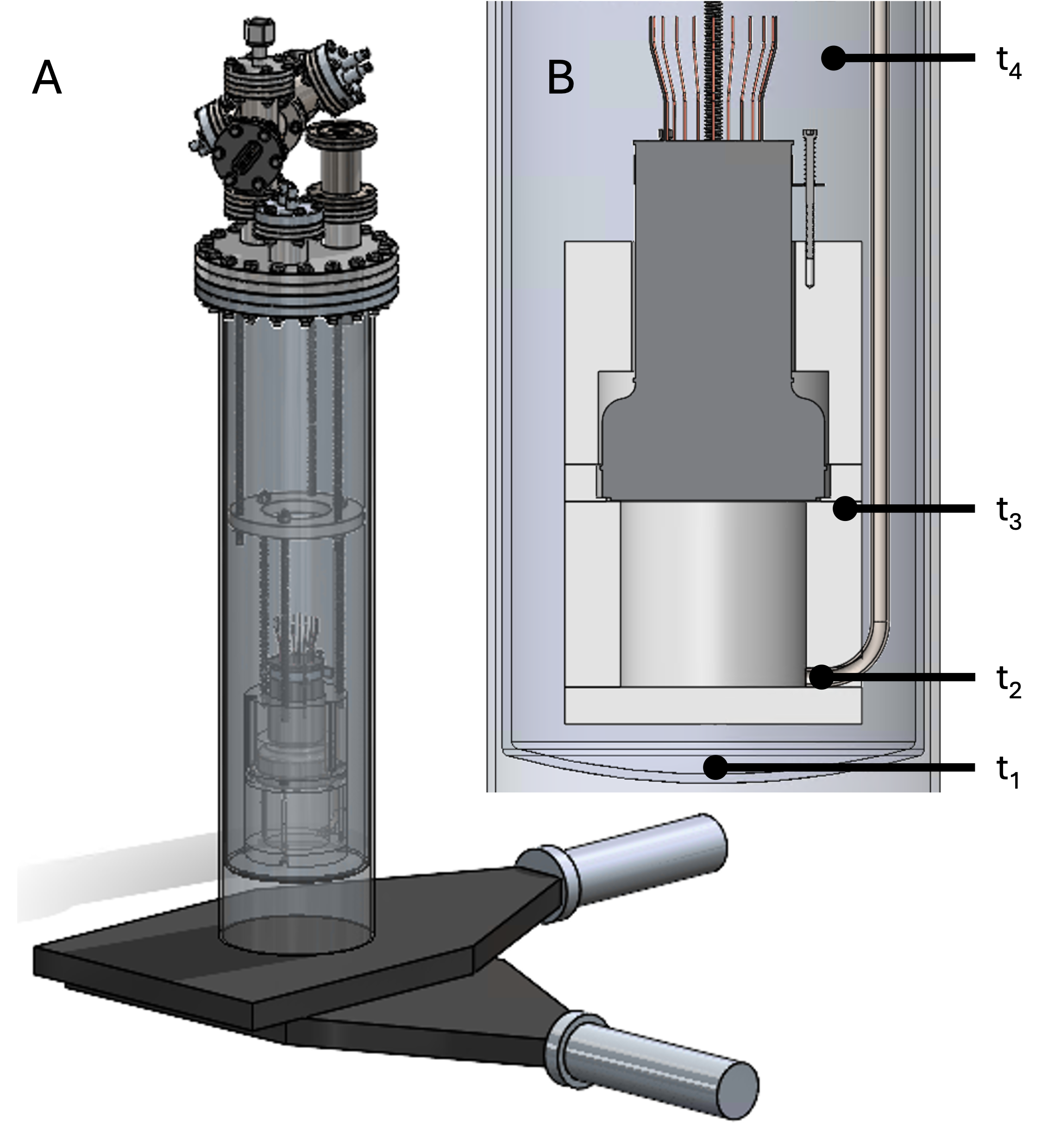}  
\caption{Overview of the detector system. Panel A shows the cryostat with the muon panels on the bottom arranged one perpendicular to the other. Inside the cryostat there is the \PTFE\ cup coupled to the \PMT. Panel B shows a zoomed section view of the \PTFE\ cup. The positions of the temperature sensors used to monitor the filling status of the detector are also shown.}
\label{fig:nlts_detector}
\end{figure}

During the entire measurement time, the detector conditions - pressure inside the cup, gas flow rate, and temperature - were recorded along with the filling status. 
As shown in Fig.~\ref{fig:nlts_detector}, this was accomplished via different RTD sensors positioned at the bottom of the cryostat ($t_1$), the bottom of the \PTFE\ cup ($t_2$), the \PMT\ level ($t_3$), and \SI{1.}{in} above the \PMT\ voltage divider ($t_4$).

As part of the detector system, two muon panels (plastic scintillator panels each coupled to a \PMT) were positioned at the bottom of the cryostat,  oriented perpendicular to each other. While the available space around the cryostat limited the size and positioning of the muon panels, they did provide enough coverage to tag the majority of muons passing through the active region.

\subsection{The Noble Liquid Test Stand}

Measurements were made in November 2023 at Pacific Northwest National Laboratory (\PNNL), using the Noble Liquid Test Stand (\NLTS), a general-purpose cryogenic detector test facility.
The \NLTS\ consists of a liquid condensing vessel and a gas recirculation system. These systems are independent of the detector in order to maintain as much flexibility as possible for use with multiple test systems. The cooling system features a condenser volume, thermally coupled to a Gifford-McMahon cryocooler, AL325 (\SI{300}{\watt} at \SI{77}{\kelvin}~\cite{BFAL325}) coldhead, which is connected by a vacuum-insulated liquid inlet line to the detector. Gas from the fill bottle or recirculation system is cooled and condensed and then driven by gravity into the detector. The condensed liquid provides the cooling power for the initial cool down of the detector. Warm gas from the detector returns to the recirculation gas panel which includes a metal bellows gas recirculation pump, purification getters (not used in this study), and ports to inject radioactive sources into the gas stream. Valves allow for the gas flow to be split to pass through or around the source. The \NLTS\ features control, monitoring, and recording of pressure, temperature, and flow through a custom slow-control software package developed at \PNNL. 
In this measurement, a UHP Grade \SI{99.999}{\%} pressurized nitrogen gas bottle from Oxarc, Inc.~\cite{Oxarc} was used as the source of nitrogen. 

\subsection{Thoron source}

\begin{figure}[tb]
\centering
\includegraphics[width=\columnwidth]{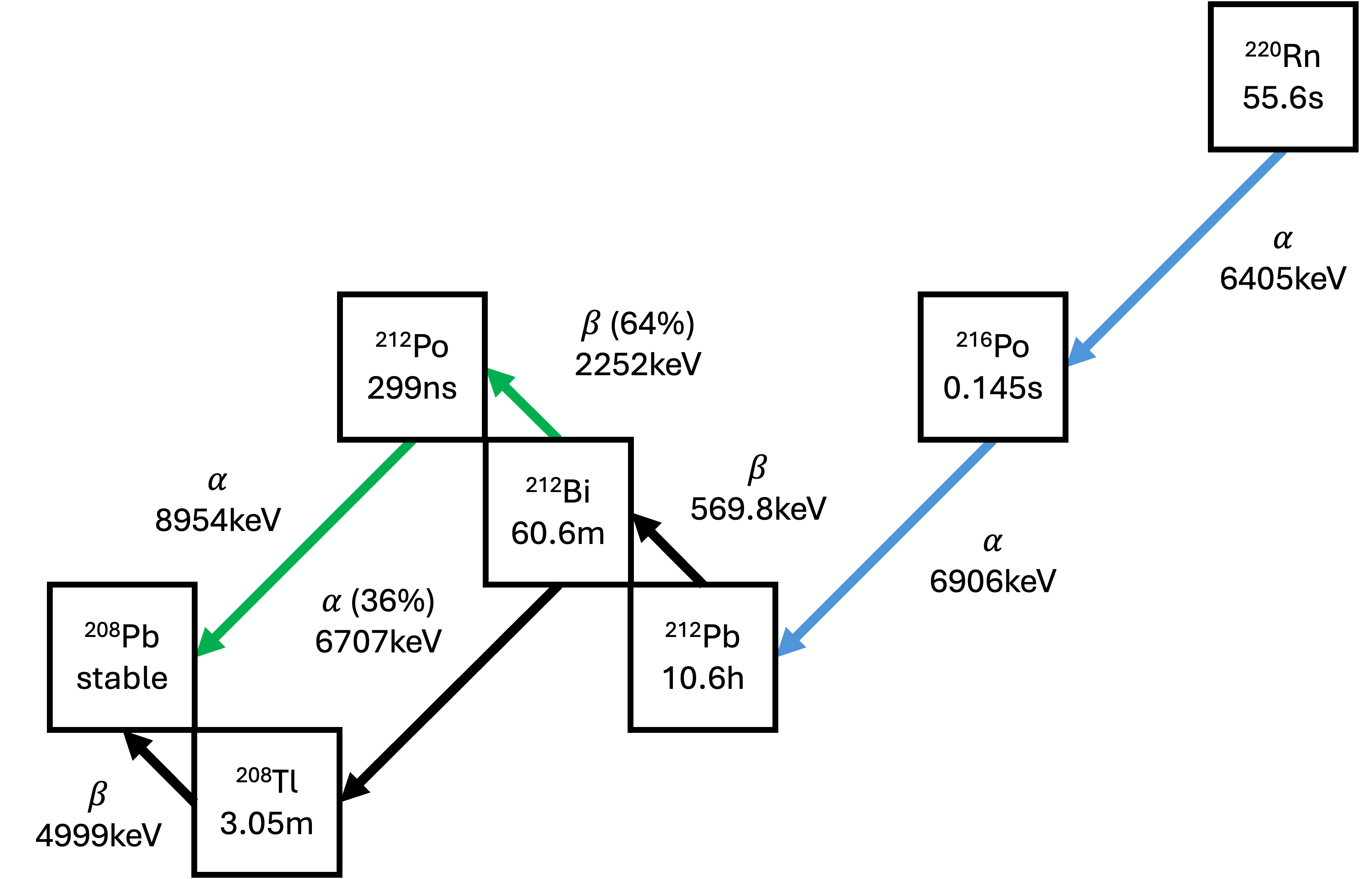}  
\caption{The decay chain of \isotope{Rn}{220} following its emanation from a \isotope{Th}{228} source~\cite{PhysRevD.95.072008, Audi_2012, NNDC}. Prompt alphas are highlighted in blue, and the \bipo\ coincidence in green.}
\label{fig:rn220decay}
\end{figure}

A commercially available flow-through \isotope{Rn}{220} (thoron) source, TH-1025 from Pylon~\cite{PylonTH1025}, was connected to the \NLTS\ gas system. The source consists of a dry powder containing \isotope{Th}{228} enclosed between glass filters. The \isotope{Th}{228}~and its immediate progeny \isotope{Ra}{224} remain contained in the source volume while \isotope{Rn}{220} is emanated and entrailed in the nitrogen gas that flows through the source. Additional vacuum compatible inline filters (replaceable sintered element with \SI{0.5}{\micro\meter} nominal pore size) were installed both upstream and downstream of the source to avoid dispersing any long-lived radioactive material. The source activity was calibrated by the manufacturer at \thoronsourceactivity.

Fig.~\ref{fig:rn220decay} shows the decay chain of \isotope{Rn}{220}.
The upper part of the chain has two alpha emissions corresponding to the decay of \isotope{Rn}{220} to \isotope{Po}{216} ($E_{\alpha}=\qty[]{6288}{\keV}$) and the subsequent decay to \isotope{Pb}{212} ($E_{\alpha}=\qty[]{6778}{\keV}$).
Because \isotope{Rn}{220} and \isotope{Po}{216} have relatively short half-lives and decay shortly after release from the source, hereafter we refer to their decay products as ``prompt alphas''.
The lower part of the chain, driven by the longer \Pbhalflifeexpected\ half-life of \isotope{Pb}{212}, includes the \isotope{Bi}{212}-\isotope{Po}{212} (\bipo) coincidence of the $\beta$-decay of \isotope{Bi}{212} ($E_{\beta}=\qty[]{2252}{\keV}$) and the subsequent $\alpha$-decay of \isotope{Po}{212} ($E_{\alpha}=\qty[]{8785}{\keV}$) with a \BiPohalflifeexpected\ half-life. 

\subsection{Electronics, Data acquisition, and Event Reconstruction}

The data acquisition (\DAQ) system is based on a CAEN V1725B, 16-channel, \SI{14}{bit}, \SI{2}{V_{pp}}, \SI{250}{\mega\hertz} digitizer~\cite{CAENV1725}. 
Signals from the \PMT\ in the nitrogen cryostat trigger the digitzer internally with a trigger threshold of \SI{\sim 24}{\milli\volt} which corresponds to roughly fifteen times the amplitude of a single photoelectron (\SPE). When triggered, the \DAQ\ system records the \PMT\ output signals from the nitrogen detector and the two muon panels over an acquisition window of \SI{3.2}{\micro\second} (\SI{0.2}{\micro\second} of pre-trigger and \SI{3.0}{\micro\second} post trigger) which is long enough to capture \SI{99.9}{\%} of the \bipo\ coincidence events in a single trigger.

The \DAQ\ software is a customized version of the \textsc{daqman} software~\cite{ben_loer_2019_3347152}, which also provided the framework for event reconstruction.
In the offline event reconstruction, a baseline for each triggered waveform was calculated using a moving average over all samples below a given threshold, and using a linear interpolation for samples above the threshold. A pulse-finding algorithm is then applied to the baseline-subtracted waveform to identify pulses based on the amplitude and shape of the waveform, from which the relevant information about the identified pulses are computed (e.g. pulse start and end times, amplitude, and integrated charge).

To account for varying \PMT\ gain in different detector conditions, the integrated charge of each pulse was normalized by the peak charge corresponding to a \SPE. The \SPE\ normalization factor was obtained on a run-by-run basis by looking for small pulses, detected by a dedicated \SPE-finding algorithm with lower threshold than the pulse-finding algorithm, in the tails of muon events (selected by requiring coincident signals in both the muon panels). The spectra of integrated charge was then fit with a Gaussian response function to extract the peak location (see Appx.~\ref{appendix:pmt} on the correction for under-amplified photoelectrons).  

\section{Data Taking}

Initial tests were performed with gas at room temperature (\RT) to assess the efficiency with which the radioactive isotopes were being introduced into the detector and to optimize the flow rate of nitrogen through the source. Just prior to cool down, the detector was filled with \avegasnitrogenpressure\ of nitrogen at \RT, and the gas was recirculated through the thoron source for \precooldowntime, to build up an initial \isotope{Pb}{212} activity within the detector (see Fig.~\ref{fig:gr_rate_BiPo_LN2}).
The cool down was initially performed with the recirculation pump on. To achieve better stability we then switched to cryo-pumping where condensation on the cold head draws nitrogen from the gas bottle. During the entire cool down and filling process the system status (temperature, pressure, and fill level) was monitored and recorded.

The liquid level reached the bottom of the cup after \daysbottomofcup\ and reached the top of the cup after \dayspmtlevel.
After an additional day of filling, where the liquid level covered the entire body of the \PMT\ (ensuring a sufficiently high buffer of liquid to keep the cup completely immersed at all times), the source valves were closed, isolating it from the gas recirculation loop. The entire filling procedure took roughly \cooldowntime, with an average gas fill rate of \fillingrate\ for a total volume of \totalfill. The detector remained full for a week in stable conditions at thermal equilibrium: \aveliquidnitrogentemperature, and \aveliquidnitrogenpressure ~\cite{NISTWebBook}.

The \PMT\ and \DAQ\ system were on and acquiring data during the cooldown, filling with liquid, and subsequent decay of source-related activity after the source valves were closed, for a total of \totdays\ of data taking. The data were then divided into different periods based on temperature, liquid fill level, and source activity in the detector, which was critical to isolating signals from the \liquidn.

\section{Analysis} 
\label{sec:analysis}

\subsection{Analysis Plan}

The first step in the analysis involves confirming that the detected signals are from the thoron source and not ambient background. In the decay chain of \isotope{Rn}{220}, the detection of the prompt alphas is hindered by the freezing of radon (freezing point at \SI{202}{\kelvin}) at \liquidn\ temperatures on the internal surfaces of the condenser, preventing it from reaching the detector. Consequently, the observed activity is predominantly governed by the relatively longer half-life of \isotope{Pb}{212}. Identifying the parent \isotope{Pb}{212} decays confirms that isotopes related to the source are reaching the detector.

The second step consists of identifying \bipo\ coincidences in the decay chain so that we can isolate the alpha decay. This is achieved by selecting events with at least two pulses (the first pulse is assumed to be the \isotope{Bi}{212}$\to$ \isotope{Po}{212} $\beta$-decay, and the second the \isotope{Po}{212} $\to$ \isotope{Pb}{208} $\alpha$-decay) and then verifying the unique time signature between them.

Finally, we need to verify that the alpha signal is originating from interactions in the liquid and not nitrogen gas or the walls of the container. We therefore performed the \bipo~search in data taken when the detector is filled with \RT\ and cold \gasn, \liquidn, and in vacuum. Comparing data taken \textit{during} filling allows one to relate the appearance of features in the spectrum to the presence of liquid inside the cup. Comparison to vacuum data allows decoupling the energy spectrum of alphas interacting in the liquid from possible radioluminescence on the detector \PTFE\ or fused silica surfaces.

\subsubsection{Variables and cuts}

In what follows, only events happening in the detector with no associated activity in the veto panels are considered. Requiring such an anti-coincidence removes the majority of events with muons passing through the detector (with an estimated rate of \muonrate\ at \PNNL~\cite{AGUAYO2012}).

To select \bipo~coincidences, only events having at least two pulses are considered. In these events the integrated charge of the triggering pulse (pulse start time at roughly \SI{0}{\micro\second} relative to the trigger) is labeled $Q_0$, while the charge of the subsequent pulse is labeled $Q_1$ (we ignore all other pulses in the trigger window). $Q_0$ is assumed to be from the \isotope{Bi}{212} $\beta$-decay, while $Q_1$ is attributed to the \isotope{Po}{212} $\alpha$-decay. Events with pre-trigger pulses (with a pulse start time \SI{<-0.05}{\micro\second}) are rejected in order to avoid mis-identifying $Q_0$ and $Q_1$. Furthermore, events where the second pulse is likely an afterpulse, defined as a pulse with a start time in the range \SIrange{0.40}{0.65}{\micro\second}, are also discarded~\footnote{Given the applied voltage and the \PMT's characteristics, afterpulses from imperfect vacuum in the PMT are expected to appear at $\delta t_{ap} \simeq \qty[]{0.258}{\micro\second}~\sqrt{\frac{M}{Q}}$ following the main pulse, where ($M/Q$) is the mass/charge ratio \cite{phd_dmp2017}. Our data indicate that our afterpulses are primarily due to a small He contamination (for He$^+$, $\delta t_{ap} \simeq \qty[]{0.516}{\micro\second}$).}.

Both $Q_0$ and $Q_1$ are scaled by the mean \SPE\ charge so that the estimated energy is expressed in photoelectrons (\PE).
The average relative error on the run-by-run \SPE\ calibration is included as a systematic (\spesystematicln, \spesystematicgn, and \spesystematicgnRT\ respectively for \liquidn, and cold and \RT\ gas).
The time difference ($\Delta t$) between the first two pulses is also calculated. 

\begin{figure}[tb]
\centering
\includegraphics[width=\columnwidth]{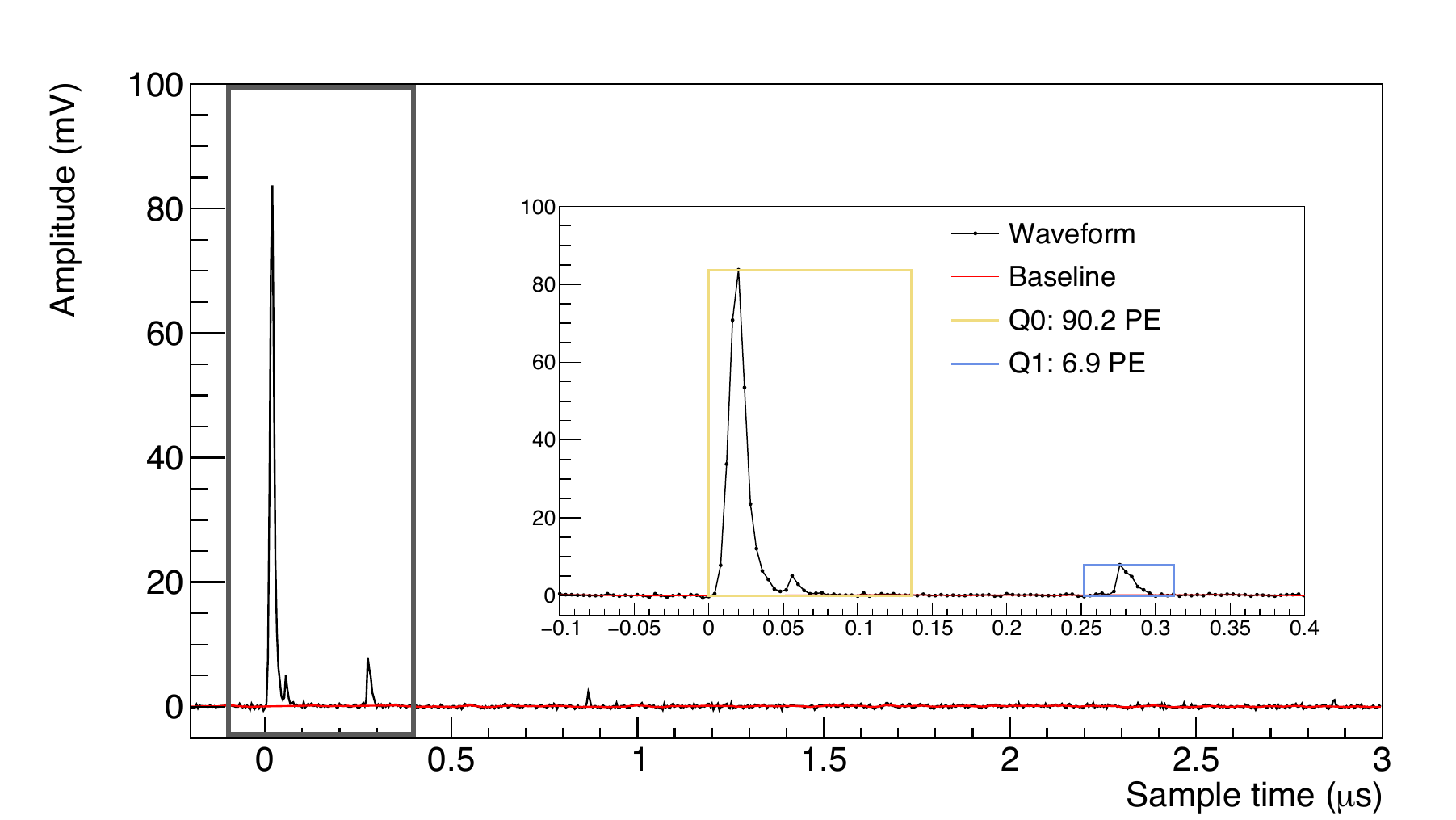}  
\caption{Example of a \bipo\ candidate event showing the full acquired waveform (black) and the calculated baseline (red). The inset shows a zoomed in view of the identified $Q_0$ and $Q_1$ pulses, marked with colored boxes. The other small pulses in the waveform are consistent with single photoelectrons.}
\label{fig:BiPo_wfm}
\end{figure}

\bipo\ candidates are selected based on a combination of cuts applied to $Q_0$ and $\Delta t$. Events with $Q_0<225$\,\PE\ are selected, where the upper threshold corresponds to the feature resembling a $\beta$-decay end-point seen in the $Q_0$ spectrum. Additionally, only events with $\Delta t$ in the range \SIrange{0.16}{1.5}{\micro\second} are considered, where the lower limit is set to avoid overlap of the two pulses and the upper limit was chosen to be $\sim 5$ times the half-life of \isotope{Po}{212}. The overall efficiency of the cuts on $\Delta t$ is \SI{48.5}{\%}. An example of an identified \bipo\ candidate is shown in Figure~\ref{fig:BiPo_wfm}.

In what follows, exponential decays are fitted with the following functional form $f(t)=a_0+a_1~e^{-\nicefrac{t}{\tau}}$, where $\tau$ is the decay constant.

\subsection{Identifying the parent \isotope{Pb}{212}}

\begin{figure}[tb]
\centering
\includegraphics[width=\columnwidth]{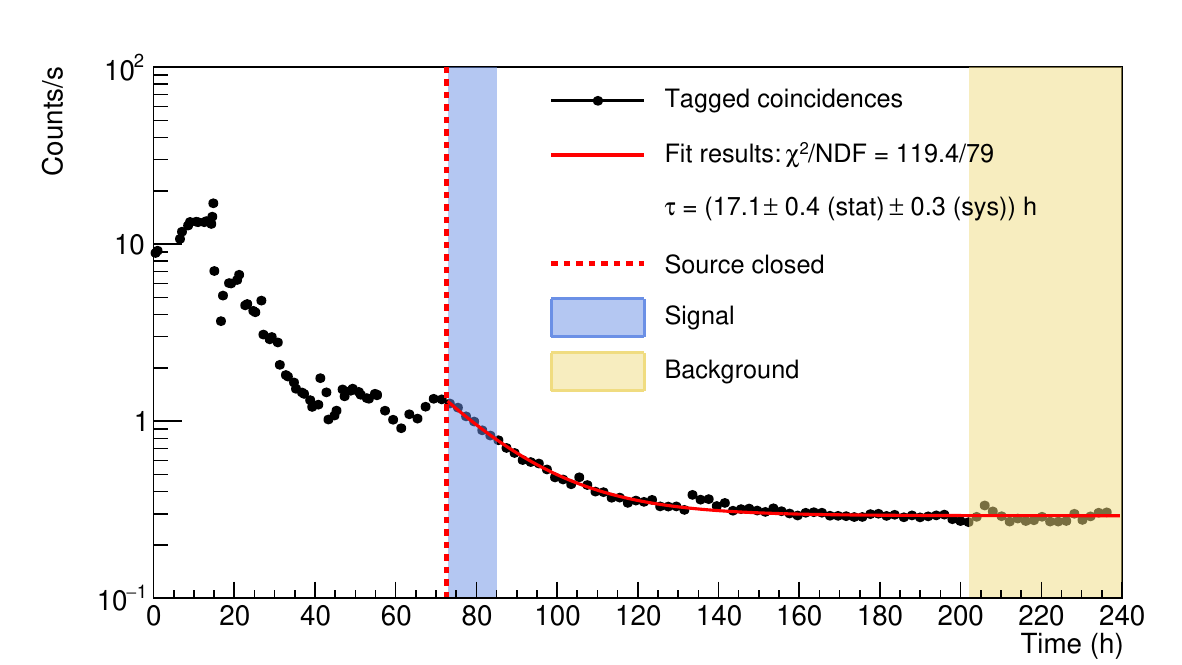}  
\caption{Tagged coincidence event rate as a function of time since the beginning of the filling process. The source was closed at \SI{72.6}{\hour} with the detector full of liquid. The ``signal'' period is defined as \SIrange{73}{85}{\hour}, while the ``background'' period is defined as \SIrange{202}{240}{\hour}.}
\label{fig:gr_rate_BiPo_LN2}
\end{figure}

Once the cup was fully filled with liquid and enough activity was built up inside the detector, the source was isolated from the system. The event rate of candidate \bipo\ coincidences as a function of time was monitored and then fit with an exponential decay (see Fig.~\ref{fig:gr_rate_BiPo_LN2}) that yielded a decay time constant of \Pbtaumeasured, close to the expected \Pbtauexpected\ mean lifetime of \isotope{Pb}{212}, indicating that these events are related to the thoron source. The discrepancy between the measured and expected decay time constant is likely due to a small flow of \isotope{Pb}{212} into the active detector (even after the source valves are closed) from the condenser region where some fraction is frozen.

The time evolution of the tagged coincidence rate also served to define the signal and background datasets (refer to Fig.~\ref{fig:gr_rate_BiPo_LN2}). In what follows, the \liquidn\ ``signal'' dataset refers to the first \SI{12}{\hour} after closing the source, while the ``background'' dataset encompasses the last \SI{38}{\hour} before warming up the detector (starting more than eight \isotope{Pb}{212} half-lifes after closing the source), when the rate of tagged coincidences reached a constant baseline value.

\subsection{Bi-Po coincidences}

The presence of \bipo\ decays in the tagged coincidence events was confirmed by analyzing the $\Delta t$ distribution between the first and second pulses in the tagged events from the ``signal'' dataset. Afterpulses and other correlated backgrounds were suppressed by subtracting the $\Delta t$ distribution of the ``background'' dataset and excluding a \dtexcludedregionPb-wide afterpulse region around \SI{0.58}{\micro\second} when fitting. The spectra were normalized by the number of triggers recorded because the afterpulse probability is related to the pulse rate and not to the elapsed time.

\begin{figure}[tb]
\centering
\includegraphics[width=\columnwidth]{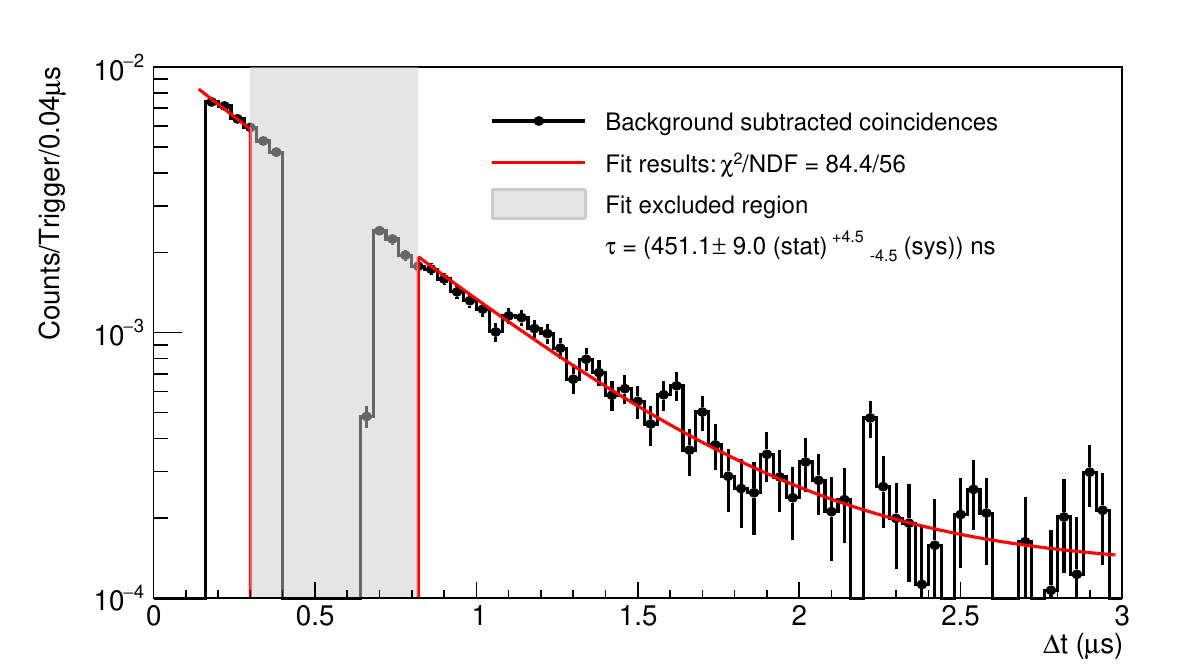}  
\caption{$\Delta t$ distribution for tagged coincidence events in the liquid signal dataset, after subtracting out the distribution from the background dataset. The residual distribution is fit with an exponential decay yielding a decay time compatible with the \bipo\ coincidence (\BiPotauexpected).}
\label{fig:dt_BiPo_LN2}
\end{figure}

The fit with a decaying exponential (see Fig.~\ref{fig:dt_BiPo_LN2}) yields a decay constant of \BiPotaumeasured, where the systematic error includes the uncertainties due to background subtraction and the effect of varying the fit range and the afterpulse-exclusion window. This is compatible with the mean lifetime of \isotope{Po}{212} (\BiPotauexpected).

\begin{figure}[tb]
\centering
\includegraphics[width=\columnwidth]{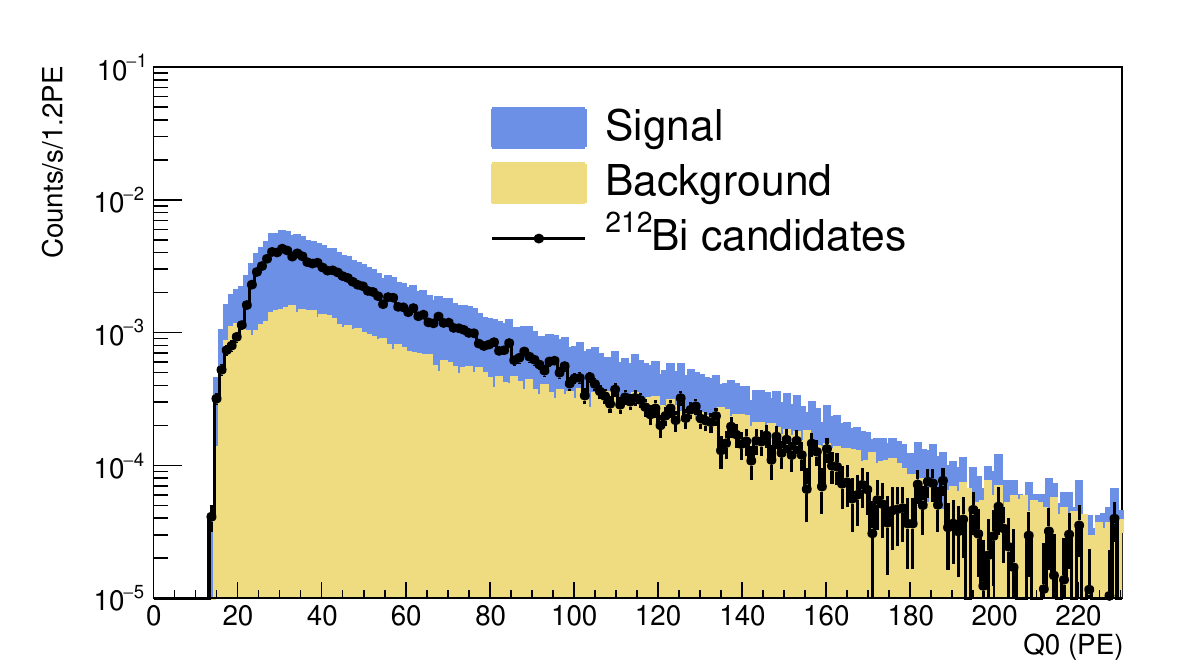}  
\caption{Comparison of the $Q_0$ energy spectrum of \bipo\ candidate events during the \liquidn~signal and background datasets.}
\label{fig:spectrum_Q0_bkgsub_ln2}
\end{figure}

\begin{figure}[tb]
\centering
\includegraphics[width=\columnwidth]{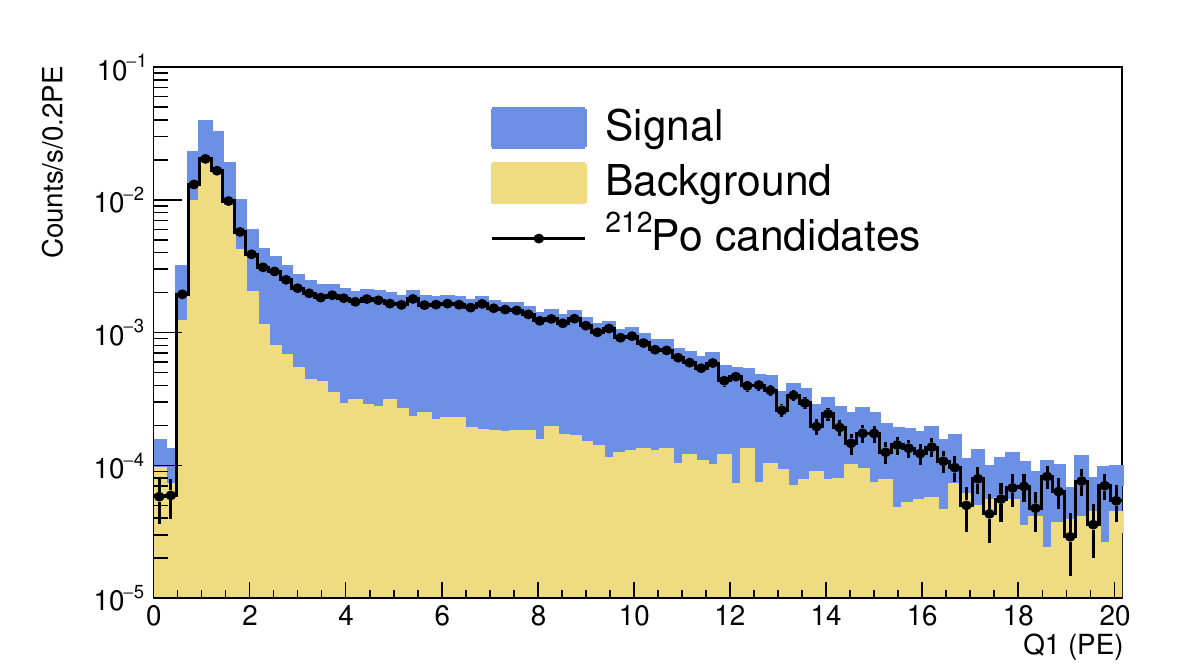}  
\caption{Comparison of the $Q_1$ energy spectrum of \bipo\ candidate events during the \liquidn~signal and background datasets.}
\label{fig:spectrum_Q1_bkgsub_ln2}
\end{figure}

Having confirmed that the events identified are primarily \bipo\ coincidences, we turn our attention to the amount of light they produce. A comparison of $Q_0$ (candidate \isotope{Bi}{212} decay) energy spectra for \bipo\ events in the ``signal'' and ``background'' datasets reveals a $\beta$-like spectrum extending up to \SI{200}{PE} (see Fig.~\ref{fig:spectrum_Q0_bkgsub_ln2}). The same spectral comparison for $Q_1$ (candidate \isotope{Po}{212} decay) reveals an excess of events from \SIrange{2}{14}{\PE} (see Fig.~\ref{fig:spectrum_Q1_bkgsub_ln2}).

Before attempting to quantify the amount of scintillation produced, we first discuss the measurements made to confirm that the signals are coming from \isotope{Po}{212} interactions in the liquid.

\subsection{Liquid vs Gas}
\label{sec:liquid_vs_gas}
The acquisition of data during the filling of the detector allows us to study \bipo\ coincidence events (same selection cuts) as the nitrogen in the cup transitioned from \RT\ to cold \gasn, and then to \liquidn, and obtain a relative measurement of the scintillation yield of \liquidn\ with respect to \gasn.

\begin{figure}[tb]
\centering
\includegraphics[width=\columnwidth]{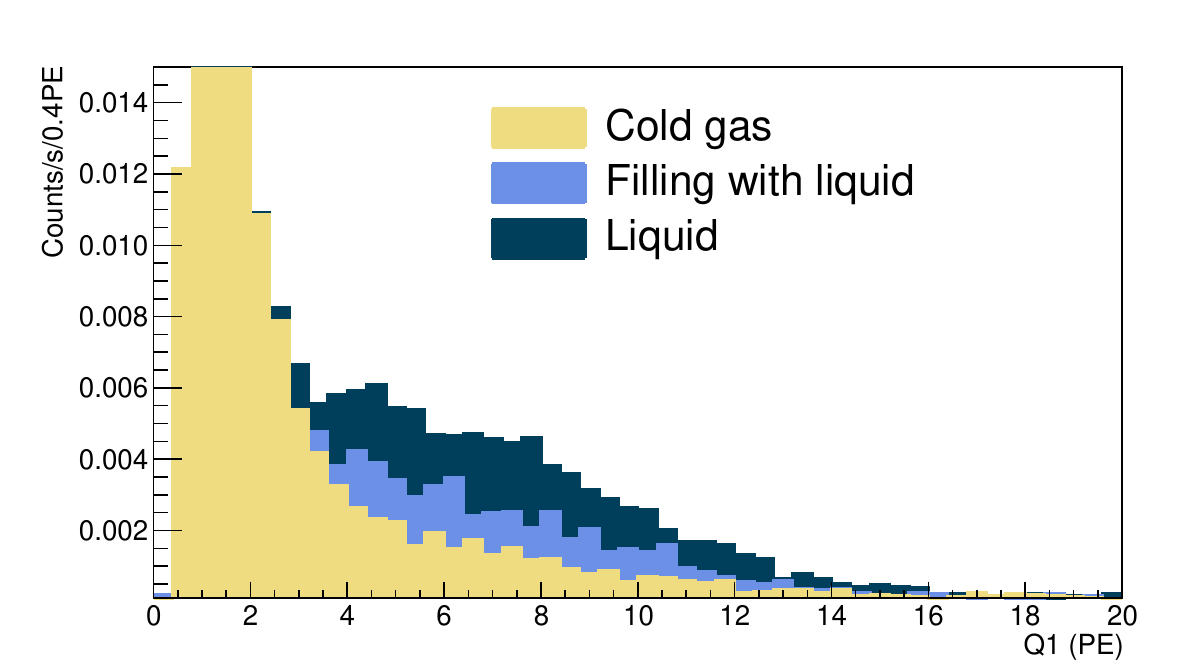}  
\caption{Evolution of the $Q_1$ spectra for \bipo\ candidates as the nitrogen in the cup transitioned from gas to liquid. A broad peak becomes more prominent as the liquid level in the cup rises.}
\label{fig:LN2_campaign_cup_filling}
\end{figure}

Fig.~\ref{fig:LN2_campaign_cup_filling} shows the $Q_1$ energy spectrum for tagged coincidence events during the transition from cold gas to liquid. The fit to the $\Delta t$ background-subtracted spectrum for cold \gasn\ yields a decay constant of \BiPotaumeasuredgn\ (with a similar result for \RT), confirming these events are \isotope{Po}{212}.  As can be seen, the broad peak-like feature at low energy appears in the $Q_1$ spectrum only when liquid starts to fill the cup and becomes more prominent as the liquid level rises, confirming that these events are associated with the presence of \liquidn\ in the cup.

\begin{figure}[tb]
\centering
\includegraphics[width=\columnwidth]{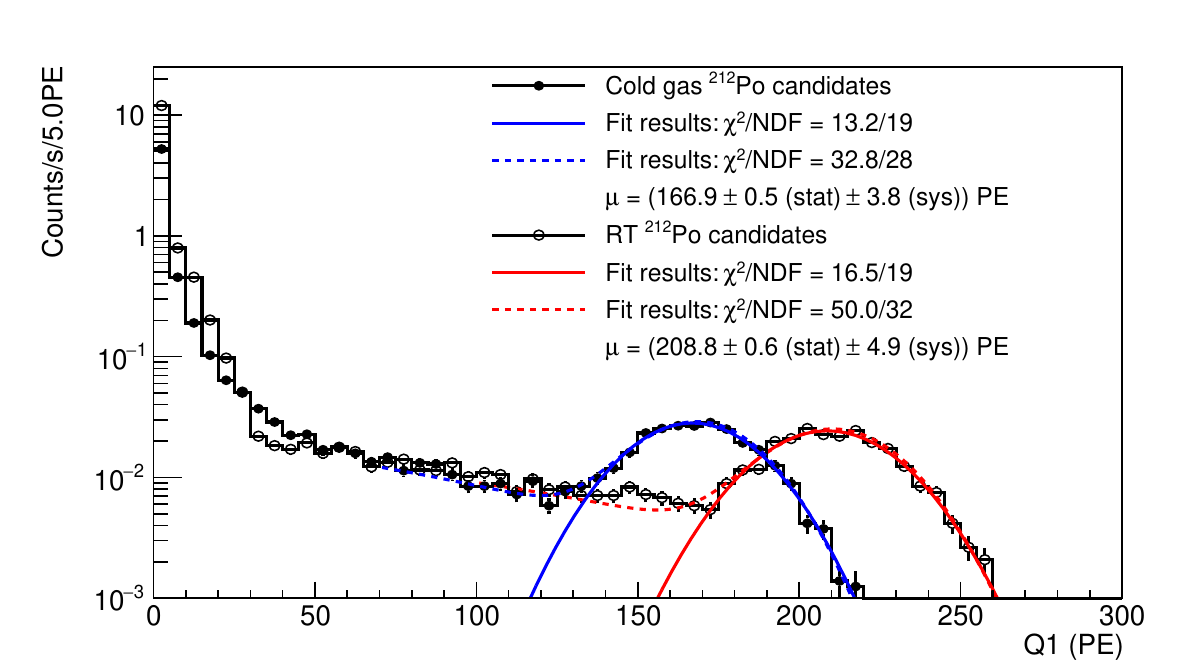}  
\caption{Energy spectrum of the \bipo\ in cold (solid markers) and \RT\ (hollow markers) gas. The spectra are fit with both a simple Gaussian (solid) and a Monte Carlo generated spectra smeared with the detector resolution (dashed). The difference between the two fitting methods is treated as a source of systematic error, as is the range over which the fits are performed.}
\label{fig:fit_Q1_Cold_and_RTGas}
\end{figure}

To evaluate the scintillation signal in gas, we analyzed data at \RT, and in cold gas just prior to the start of condensation of nitrogen in the cup. Fig.~\ref{fig:fit_Q1_Cold_and_RTGas} shows the \isotope{Po}{212} alpha $Q_1$ spectrum for \bipo~candidates from a \SI{2}{\hour} run, where the averaged pressure and temperature of the gas inside the cup were \avepressuregn\ and \avetemperaturegn\ for cold gas (\avepressuregnRT, and \avetemperaturegnRT\ for \RT) - here the higher variance is due to the on-going process of filling. 

\subsection{Geant4 Simulation}
\label{sec:mc}

We employed a Monte Carlo energy-transport simulation to understand the signal shape for the \isotope{Po}{212} alpha decay in cold and \RT\ \gasn. The simulation uses the 
\textsc{geant4}~\cite{AGOSTINELLI2003250, Allison2006, ALLISON2016186} toolkit, version 10.7.p03. 
We simulate a generic \textsc{geant4} volume that contains only the active detector region with a simplified geometry consisting of a \PTFE\ cup filled with either \liquidn\ or \gasn, topped with a fused silica disk (\SI{2}{\milli\metre} thick, \SI{76}{\milli\metre} OD) representing the \PMT\ window. Details such as the external cryostat, screws, fill line, etc. are omitted. In all simulations, the primary particles are distributed uniformly within the internal volume of the cup, and the total energy deposited in the liquid or gas is recorded.

\isotope{Po}{212} alphas were simulated in \gasn\ at cold ($T = \qty{240}{\kelvin}$, $\rho = \qty{2.651e-3}{\gram\per\centi\metre\cubed}$) and \RT\ \gasn\ ($T =\qty{297}{\kelvin}$, $\rho = \qty{1.988e-3}{\gram\per\centi\metre\cubed}$). A low-energy tail is expected in the alpha energy spectrum as some alpha particles only partially deposit their energy in the gas due to the limited size of the cup relative to their range~\footnote{At the operational pressure, the \isotope{Po}{212} alpha range in nitrogen gas varies from \SI{52.9}{\mm} at \RT\ to \SI{41.4}{\mm} at \SI{240}{\kelvin}, and is \SI{0.135}{\mm} in liquid.}. To fit the simulated energy distribution to the experimental data, the spectrum is converted to \PE\ by applying a linear scaling to account for the light collection of the cup and the detection efficiency of the \PMT, and then convolved with a Gaussian resolution function (assumed constant over the energy range). 
Fitting the \isotope{Po}{212} alpha spectra in gas (Fig.~\ref{fig:fit_Q1_Cold_and_RTGas}) with the Monte Carlo generated spectrum yields a full-energy peak value of \mcfitgn\ and \mcfitgnRT\ for cold and \RT\ gas, respectively. This is consistent with the values obtained from a pure Gaussian fit to the peak region. Our final estimate for the peak value is $\mu_{Q_1}^{\gasn}$ at \qonemugn\ for cold gas, and \qonemugnRT\ for \RT, where the systematic error includes the average relative error on the \SPE\ calibration - \spesystematicgn\ (\spesystematicgnRT\ for \RT), the sensitivity of the fit to the fitting range - \fitrangesystematicgn\ (\fitrangesystematicgnRT), and the variation between using a Gaussian, or the smeared Monte Carlo PDF - \fitstratsystematicgn\ (\fitstratsystematicgnRT). The reduced scintillation yield in cold gas compared to room temperature is expected due to the increased collisional quenching in the denser cold gas~\cite{AVE200741, AVE200850} and corrections for the temperature and pressure dependence of \gasn\ scintillation are discussed later in Section~\ref{sec:relative_yield}.

\isotope{Po}{212} alpha decays were also simulated in liquid to look for possible alternate mechanisms for the production of Cherenkov light from alpha decays (see Appx.~\ref{appendix:relative_quenching} for more details). Alpha particles from \isotope{Po}{212} decays cannot produce Cherenkov emission directly because they are well below the minimum energy threshold, but they can produce gammas from inelastic interactions with the nitrogen nuclei. These gammas can then Compton scatter, producing sufficiently high energy electrons to produce Cherenkov light. In our simulations, Cherenkov light is only produced in \SI{1.7e-3}{\percent} of \isotope{Po}{212} decays (in agreement with similar studies in water~\cite{ackerman2012potential}), and we therefore did not include this contribution in our analysis.

\subsection{Liquid vs Container}

The relatively long lived \isotope{Pb}{212} can build up and settle on the internal surfaces of the detector, producing alpha interactions not just in the liquid but also in the \PTFE\ of the detector walls or the \PMT's fused silica window. These interactions could potentially produce radioluminescence that could be mistaken for scintillation.

\begin{figure}[tb]
\centering
\includegraphics[width=\columnwidth]{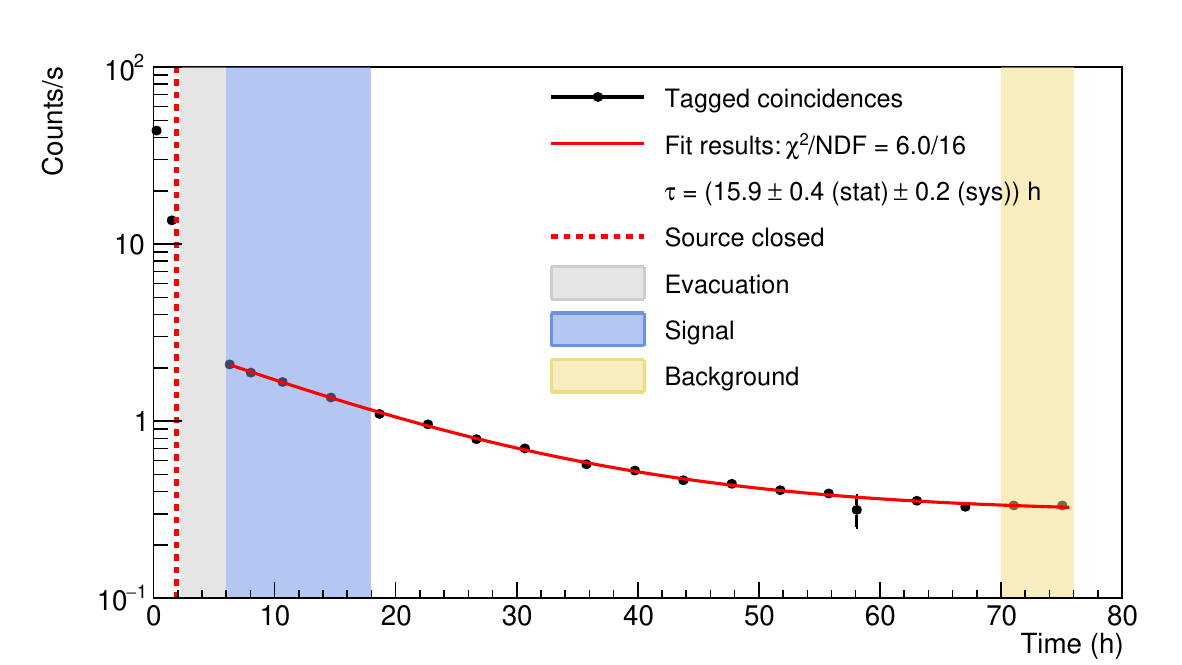}  
\caption{\bipo\ event rate as a function of time since the beginning of the vacuum data set. Gaseous nitrogen passing through the source was recirculated through the detector until the source was closed at $t = \qty[]{1.9}{\hour}$, following which the detector was evacuated (no data acquired during pump down). The ``signal'' dataset is defined from \SIrange{6}{18}{\hour} while the ``background'' dataset is defined from \SIrange{70}{76}{\hour}.}
\label{fig:gr_rate_BiPo_V}
\end{figure}

We looked for radioluminescence in the same setup with a second data taking campaign in the absence of nitrogen. The detector was initially filled with \gasn\ at \avepressuregnvacuum\ and the gas was continuously recirculated through the thoron source. Once enough activity was built up in the active volume of the detector, the chamber was evacuated, during which period the PMT was turned off to avoid high voltage discharges on the \PMT\ base at intermediate vacuum levels. Once the pressure dropped below \avepressurevacuum\ the \PMT\ was turned back on and data were collected for \totdaysvacuum. The first \SI{12}{\hour} of vacuum data are considered ``signal'', whereas the last \SI{6}{\hour} are classified as ``background'' (see Fig.~\ref{fig:gr_rate_BiPo_V}). The same analysis (data selection and cuts) described previously for \gasn\ and \liquidn\ was performed on the vacuum data.

Fitting the event rate as a function of time after the source was closed and the detector evacuated (see Fig.~\ref{fig:gr_rate_BiPo_V}) yields a decay time constant of \Pbtaumeasuredvacuum\footnote{This is in good agreement with the \SI{15.3}{\hour} mean lifetime of \isotope{Pb}{212}, as one would expect for the vacuum runs where there is no additional source of \isotope{Pb}{212} from the condenser or recirculation systems.}, while the fit of the $\Delta t$ spectrum yields \BiPotaumeasuredvacuum, confirming the presence of \bipo\ radioluminescence signals. 

\begin{figure}[tb]
\centering
\includegraphics[width=\columnwidth]{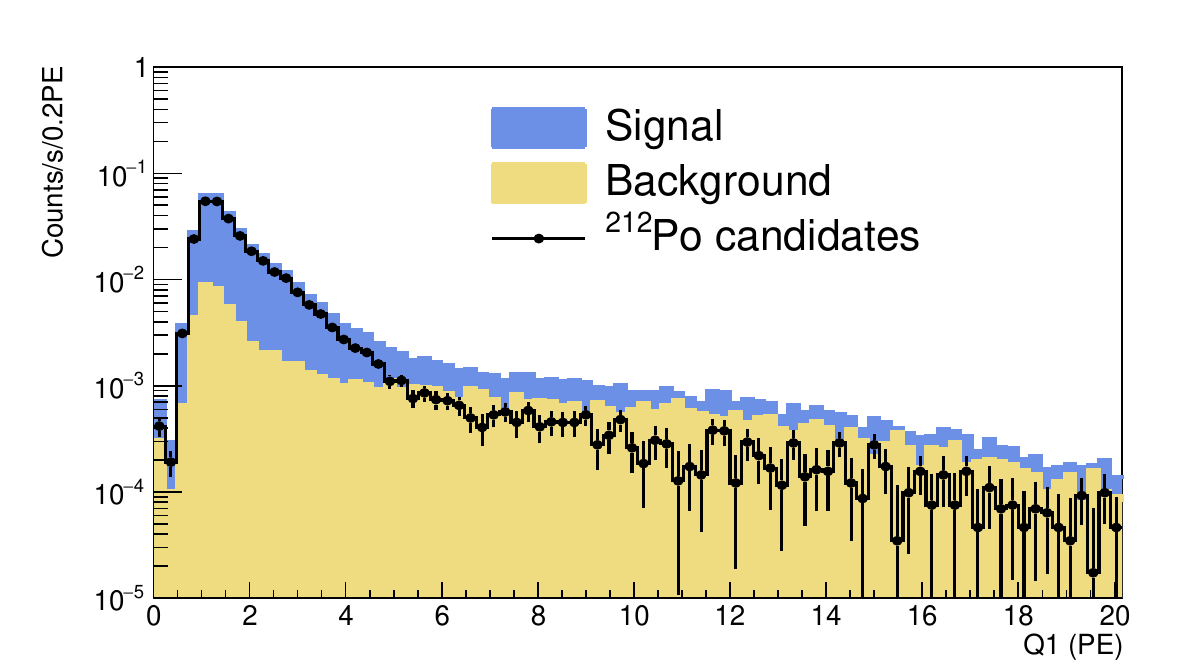}
\caption{Energy spectra of tagged \isotope{Po}{212} events in vacuum. The background-subtracted spectrum ($S_V(Q_1)$) is shown in black.}
\label{fig:spectrum_Q1_bkgsub_v}
\end{figure}

Fig.~\ref{fig:spectrum_Q1_bkgsub_v} shows the background subtracted spectrum of the \isotope{Po}{212} radioluminescence events from the evacuated detector. These events follow an exponentially falling spectrum, with the majority of events below \SI{6}{\PE}. Note that these events should also be present in the gas and liquid nitrogen datasets (for example in Fig.~\ref{fig:LN2_campaign_cup_filling}) and in the next section we use this background subtracted energy spectrum of the \isotope{Po}{212} ($S_V(Q_1)$) as the reference PDF for radioluminescence events.

\section{Results}

\subsection{Energy spectrum and fit}
\begin{figure}[tb]
\centering
\includegraphics[width=\columnwidth]{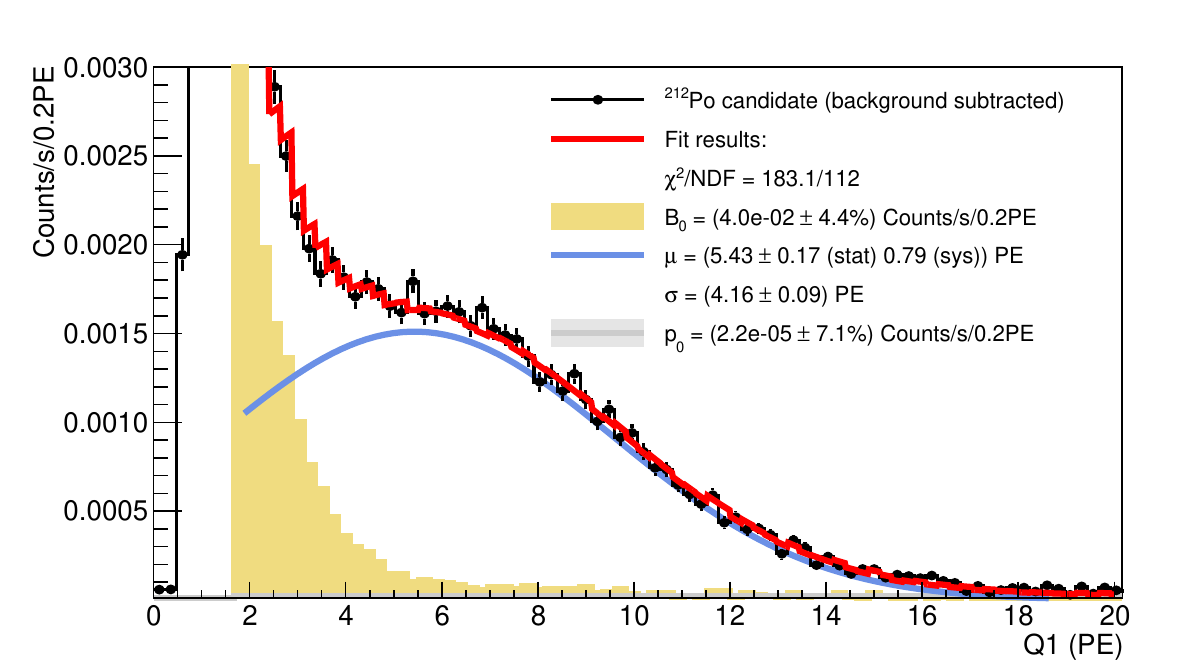}  
\caption{Fit of the background-subtracted $Q_1$ energy spectrum of Figure~\ref{fig:spectrum_Q1_bkgsub_ln2} with $f(Q_1)$. The scintillation light resulting from the \SI{8785}{\keV} alpha is modeled as a Gaussian (blue) on top of the radioluminescence spectrum (yellow) and a flat distribution (gray) representing any residual, unaccounted backgrounds.}
\label{fig:fit_Q1}
\end{figure}

To quantify the scintillation signal in liquid we fit the background-subtracted energy spectrum shown in Fig.~\ref{fig:spectrum_Q1_bkgsub_ln2}. We assumed a Gaussian distribution for the \isotope{Po}{212} alpha spectrum ($S_{\alpha}(Q_1) = A_0 e^{-\nicefrac{\left(Q_1-\mu\right)^2}{2 \sigma^2}}$), and used the radioluminescence PDF obtained from the vacuum run ($S_V(Q_1)$) with a free normalization factor $B_0$. We found that a small flat component ($p_0$) was also needed to match the data above \SI{\sim 15}{\PE}. 
The overall function takes the form $f(Q_1) = B_0 S_V(Q_1) + S_{\alpha}(Q_1) + p_0$. 
The fitted model shows reasonable agreement with the data (see Fig.~\ref{fig:fit_Q1}) and yields a scintillation signal of \qonemuln\ for the \SI{8785}{\keV} \isotope{Po}{212} alpha. 
The systematic error includes uncertainties in background estimation (radioluminescence - \bkgsystematicvacuum, and flat contribution - \bkgsystematicflat), the effect of changing the fitting range and the starting point of the fit (\bkgsystematicpemin), and the average relative error on the
SPE calibration (\spesystematicln).

\subsection{Relative yield to gas}
\label{sec:relative_yield}

While our experimental setup does not allow for an absolute measurement of the scintillation yield, we can compare the measured yield to that of gaseous nitrogen in the same setup. The scintillation yield of liquid nitrogen ($Y_{\liquidn}$) relative to that of gas at standard temperature and pressure $\left(Y_{\gasn}^{\STP}\right)$ is calculated as the ratio of the alpha peak positions measured in liquid and gas (cold or \RT), and then corrected for the pressure ($p$) and temperature ($T$) dependence of the \gasn\ scintillation yield:
\begin{align}
\label{eq:ratio}
    \frac{Y_{\liquidn}}{Y_{\gasn}^{\STP}} &= \frac{Y_{\liquidn}}{Y_{\gasn}^{(p,T)}} \cdot \Theta(p,T) \nonumber \\
    &= \frac{\nicefrac{\mu_{\liquidn}}{\DE_{\liquidn}}}{\nicefrac{\mu_{\gasn}(p,T)}{\DE_{\gasn}}} \cdot \Theta(p,T)
\end{align}
where $\Theta(p,T)$ is the gas scintillation yield correction with respect to \STP, and the alpha peak positions ($\mu$) are corrected for the detection efficiency (\DE) at the corresponding scintillation wavelength ($\lambda$) and temperature. The scintillation photons \DE\ depends on the \LCE\ of the \PTFE\ cup and the quantum and collection efficiency of the \PMT:
\begin{equation}
\label{eq:de}
\DE_x \equiv \DE(\lambda_x, T_x) = \LCE(\lambda_x, T_x) \cdot \QE(\lambda_x, T_x)
\end{equation}

To calculate the ratio in Eqn.~\ref{eq:ratio} we make two assumptions. In the absence of any measurement of the spectrum of the \liquidn\ scintillation, we assume that it has roughly the same emission wavelengths as gas scintillation (primarily at $\lambda_0 = \SI{337}{\nano\meter}$). This is true for noble gas scintillators~\cite{APRILEbook}, but needs to be verified for liquid nitrogen\footnote{We note that measurements of UV-excited luminescence in liquid nitrogen observed emission in the \SIrange{380}{560}{\nano\meter} range with an exponential decay on the timescale of seconds~\cite{kirko2015luminescence}.}. We also assume that the light collection is unaffected by the change in the index of refraction (see Tab.~\ref{tab:ethparticles}), or attenuation length (assumed to be large compared to the dimensions of the \PTFE\ cup), between gas and liquid. Under these assumptions $\LCE_{\liquidn} = \LCE_{\gasn} = \LCE$ (see Appx.~\ref{appendix:relative_quenching}), and $\lambda_{\liquidn} = \lambda_{\gasn} = \lambda_0$ such that Eq.~\ref{eq:ratio} reduces to:
\begin{equation}
\frac{Y_{\liquidn}}{Y_{\gasn}^{\STP}} = \frac{\nicefrac{\mu_{\liquidn}}{\QE_{\liquidn}}}{\nicefrac{\mu_{\gasn}(p,T)}{\QE_{\gasn}}} \cdot \Theta(p,T)
\end{equation}

Using the scintillation yield correction measured by the \airfly\ collaboration~\cite{AVE200741, AVE200850}, we calculate 
\airflycorrection\ for cold gas, and \airflycorrectionRT\ at \RT. The uncertainties account for the pressure and temperature variation inside the detector and the uncertainties in the parameters of the correction function. For details on how this correction was calculated, refer to Appx.~\ref{appendix:airfly}. The \PMT\ \QE\ at the different temperatures was estimated as described in Appx.~\ref{appendix:pmt}, giving a value for the ratio $\nicefrac{\QE_{\gasn}}{\QE_{\liquidn}}$ of \QEratio\ for cold gas, and \QEratioRT\ at \RT.

The relative yield is then calculated to be \LNrelativeyield\ based on the cold gas and \LNrelativeyieldRT\ based on the \RT\ gas measurements. The agreement between these values confirms that our measurements in \gasn~agree with the expected scintillation yield variations with temperature from the \airfly\ collaboration. Hereafter, we consider the relative yield based on only the cold gas measurements.

\subsection{Absolute yield}

There have been several measurements of the absolute scintillation yield of alpha particles in gaseous nitrogen ($Y_{\gasn}^{\STP}$) from both the 1950s (summarized in Ref.~\cite{BIRKS1964570}) and more recently~\cite{ LEHAUT201557, MORII2004399} though the values are not consistent and range from roughly \SIrange{100}{200}{PH/\MeV}. Combining the measurements (following the statistics recommendations from Ref.~\cite{PDG2020}), the average yield is calculated to be \GNabsoluteyieldave.

\begin{table}[t]
    \centering
    \begin{tabular}{lc}
        \hline
        Contribution & Systematics (\%) \\
        \hline
        Data analysis & 15 \\
        $\Theta(p,T)$ (\airfly~\cite{AVE200741, AVE200850}) & 4.8 \\
        \PMT\ \QE\ ratio & 14 \\
        \gasn\ yield (from literature ~\cite{BIRKS1964570, LEHAUT201557, MORII2004399}) & 9.1 \\
        \hline
        Total & 23 \\
        \hline
    \end{tabular}
    \caption{Breakdown of contributions to the systematic error for the absolute scintillation yield of liquid nitrogen. See text for details.}
    \label{tab:ylnsystematic}
\end{table}

Using this value, we can calculate the absolute scintillation yield of alpha decays in liquid to be \LNabsoluteyield.
Tab.~\ref{tab:ylnsystematic} summarizes the individual contributions to the systematic uncertainty. 

\section{Summary and Discussion}

Our measurements demonstrate that \liquidn\ produces measurable, albeit very faint, scintillation light when exposed to high energy alphas. 
To isolate the contribution of scintillation, Cherenkov emission was excluded by using a \isotope{Rn}{220} source and selecting \isotope{Po}{212} alpha decays (below the Cherenkov threshold) with the time coincidence of \bipo\ events. Radioluminescence from alphas interacting with the internal surfaces of the detector was also accounted for by identifying \bipo\ coincidences while the detector was evacuated and subtracting their contribution. To our knowledge, this is the first demonstration of scintillation in \liquidn.

By measuring the response of our detector to alpha decays in \gasn, and assuming that the detection efficiency of the setup was the same for gas and liquid scintillation (other than the change in the \PMT's \QE, which was accounted for), we find that \liquidn\ has an alpha scintillation yield that is only \LNrelativeyieldPercentage\ that of \gasn\ at \STP. Based on the average measured absolute scintillation yield of \gasn\ in literature, the absolute alpha scintillation yield of \liquidn\ is estimated to be \LNabsoluteyieldTOT. This is significantly lower than the typical alpha scintillation yield observed in noble liquids (approximately \SIrange{35000}{70000}{PH\per\MeV}~\cite{APRILEbook}). In gaseous nitrogen not all molecules excited by the passage of ionizing radiation emit photons, as they can also transfer their energy to other molecules through collisions~\cite{AVE200741}. This process, referred to as collisional quenching, causes the scintillation yield to change with pressure and temperature, as the collision rate depends on the average molecular separation and velocity. Given the roughly $650\times$ larger density of liquid compared to gaseous nitrogen at \STP, it is possible that the high rate of collisions in liquid compared to the lifetime of the scintillation excited state is responsible for the low yield. The low scintillation yield also likely explains the lack of a detected signal in Ref.~\cite{CERNpreprint} and the absence of other measurements in literature. The relative and absolute values of the scintillation yield of liquid nitrogen calculated in this work assume the same emission spectrum as gaseous nitrogen, which needs to be verified by an independent measurement of the liquid scintillation spectrum, a difficult task given the low scintillation yield. 

While the scintillation yield is low, scintillation light, in addition to Cherenkov emission, from interactions of ambient radioactivity in liquid nitrogen is likely enough to create a significant background in the Oscura CCD dark matter experiment. To address this, the surface of the CCDs will be coated with a thin opaque layer of aluminum to prevent light from reaching the active volume of the CCDs~\cite{cervantes-vergara2023skipper}.  

\section{Acknowledgments}

This work was supported by the US Department of Energy (DOE) Office of Science, Office of High Energy Physics.  
Pacific Northwest National Laboratory (PNNL) is operated by Battelle Memorial Institute for the DOE under Contract No. DE-AC05-76RL01830. PNNL-SA-217490.

\appendix

\section{Hamamatsu R11065 \PMT} 
\label{appendix:pmt}

\begin{figure}[tb]
\centering
\includegraphics[width=\columnwidth]{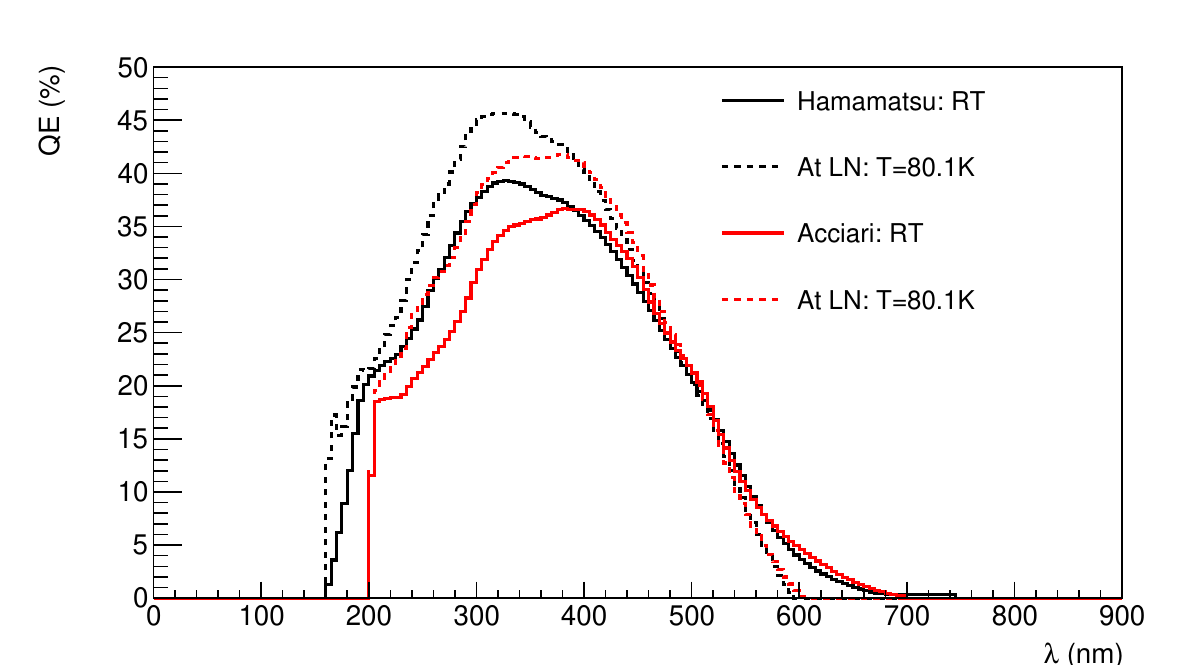}  
\caption{\PMT\ \QE\ as a function of wavelength. The solid curves are obtained by scaling the nominal \QE\ curves for this model~\cite{HR11065, RAcciarri2012} by the manufacturer measured \QE\ for our specific \PMT. Dashed lines represent the \PMT\ \QE\ extrapolated to \liquidn\ temperatures~\cite{LYASHENKO2024}. See text for details.}
\label{fig:pmt_qe}
\end{figure}

All measurements in this paper were made with a Hamamatsu R11065 \PMT\ that was originally developed for low-temperature and low-radioactivity operation in liquid argon experiments~\cite{HR11065}. There are two different models for the R11065 \QE~as a function of wavelength~\cite{HR11065, RAcciarri2012}, as shown in Fig.~\ref{fig:pmt_qe}, where the curves are scaled to match the manufacturer specified \QE\ of \SI{33.4}{\%} at \SI{420}{\nano\metre} at \RT\ for our \PMT\ (SN: BA0141). The discrepancy between the two curves (\SI{10.4}{\%} at $\lambda_0=\SI{337}{\nano\meter}$) is taken as a systematic uncertainty in the quoted \QE\ throughout the paper.

The variation of \QE\ with respect to \RT\ was evaluated based on the measurements done in Ref.~\cite{LYASHENKO2024}.
That paper reports the variation of \QE\ as a function of wavelength and temperature down to \SI{163}{\kelvin} for Hamamatsu R11410-10 \PMT s~\cite{HR11410-10}, which use a similar bialkali photocathode to the R11065 model we used. We assumed the same \QE\ change as a function of temperature (\QEgrowth\ at \SI{340}{\nano\metre}) down to \liquidn\ temperatures. With this correction, the \PMT\ \QE\ at $\lambda_0$ of \QEgasRT\ at \RT\ becomes \QEliquid, and \QEgas, at \num{80.1} and \SI{240}{\kelvin}, respectively.

For comparisons to simulations, we assumed a \SI{90}{\%} photoelectron collection efficiency and that the mean charge of a \SPE, including under-amplified signals, is $f=\qty[]{80}{\%}$ of the peak charge identified by our \SPE-finding algorithm, based on measurements of the R11410 \PMT~\cite{LUNG201232, saldanha2017model} which has the same 12-stage box and linear-focused dynode structure.

\section{Scintillation yield variation with pressure and temperature} \label{appendix:airfly}

The correction, $\Theta(p,T)$, to the scintillation yield of \gasn\ at \STP\ for different pressures $p$, and temperatures $T$, was obtained from the AIRFLY collaboration~\cite{AVE200741, AVE200850}, that studied the fluorescence yield for charged particles in nitrogen and air.
When considering pure nitrogen, it takes the form
\begin{equation}
\label{eq:airfly}
\begin{split}
\Theta(p,T) &= \frac{Y(\lambda,p,T)}{Y_0(\lambda_0,p_0,T_0)~I(p_0, T_0)} \\
            &= \frac{1+\frac{p_0}{p^{'}}}{1+\frac{p}{p^{'}} \left(\sqrt{\frac{T}{T_0}} ~ h(T_0,T)\right)^{-1}}
\end{split}
\end{equation}
where $Y$ is the yield of an emission band at wavelength $\lambda$, $Y(\lambda_0,p_0,T_0)$ is the absolute reference yield at \STP\ ($p_0=\qty[]{760}{Torr}$, and $T_0=\qty[]{273.15}{\kelvin}$) of the $\lambda_0 =\SI{337}{\nano\meter}$ band, $I(p_0, T_0)$ is the relative intensity of the $\lambda$ band with respect to $\lambda_0$, \pprimeairfly\ is the collisional quenching reference pressure~\cite{AVE200741, NAGANO2003293}, and $h(T,T_0) = \left(\nicefrac{T}{T_0}\right)^{\alpha}$ is the temperature dependence due to collisional quenching with \alphaairfly~\cite{AVE200850}.
Given the sensitivity range of the \PMT\ used in this experiment (see Appx.~\ref{appendix:pmt}) and the emission bands of \gasn, only the primary emission at $\lambda_0$ is considered for the central value of the correction (i.e., $I(p_0, T_0) = \qty[]{100}{\%}$). The correction variation depending on the choice of wavelength bands was estimated based on the emission data in Refs.~\cite{AVE200741, AVE200850} and found to be \ThetaTpcorrectionbandvariation, which is included in the uncertainty on the correction.

\begin{figure}
\centering
\subfloat[Measured scintillation yield of gaseous nitrogen, corrected for temperature and pressure, relative to the average value at \RT\ and \SI{1315}{Torr}. The roughly constant value across the entire range of operating parameters validates the correction function used.]{\includegraphics[width=\columnwidth]{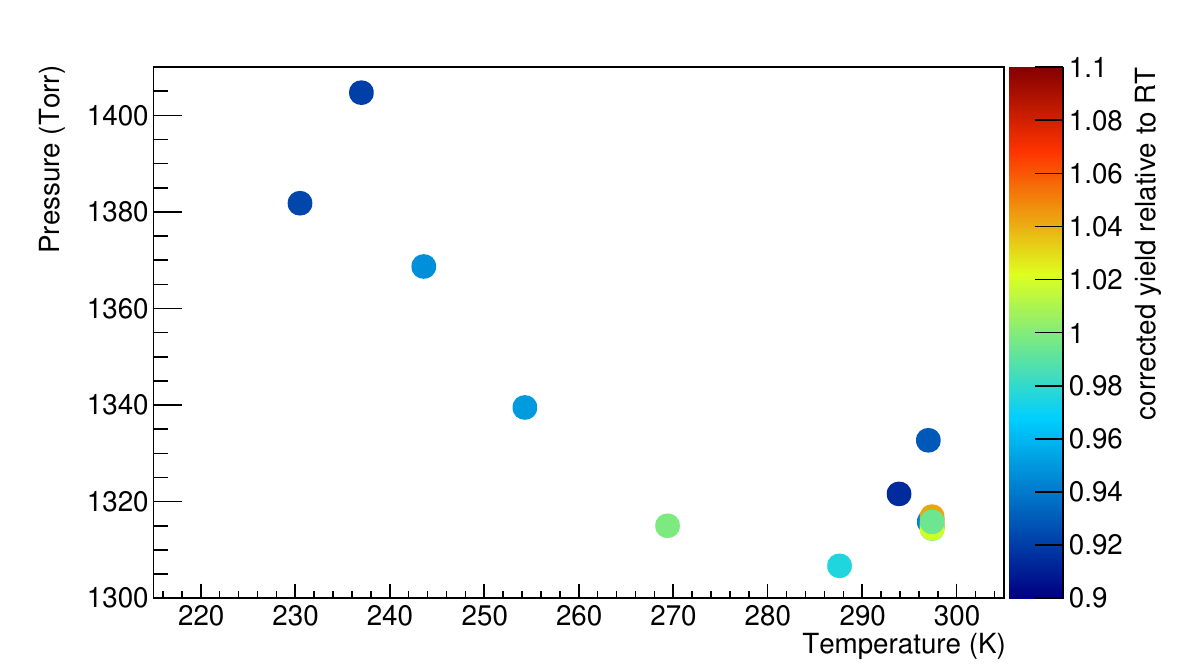} \label{fig:airfly_crosscheck_surface}} \\
\subfloat[Data (with errors) as a function only of the temperature inside the cup, showing the constant trend.]{\includegraphics[width=\columnwidth]{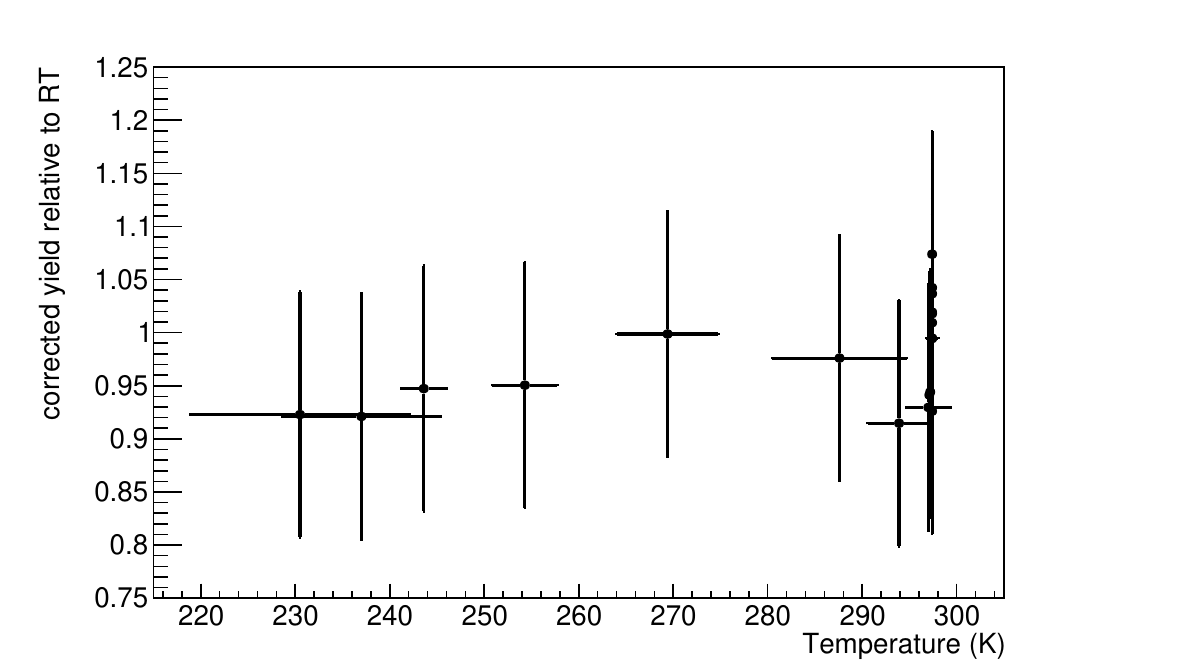} \label{fig:airfly_crosscheck}}
\caption{Cross-checks summary for the applicability of the \airfly\ correction to the data presented here.} \label{fig:airfly_summary}
\end{figure}

When Eq.~\ref{eq:airfly} is re-arranged as follows
\begin{equation}
\label{eq:airfly_re-arranged}
Y_0(\lambda_0,p_0,T_0) = \frac{Y(\lambda,p,T)}{\Theta(p,T)}
\end{equation}
it is possible to obtain an additional cross-check for the validity of the correction (see previous text in Sec.~\ref{sec:relative_yield}).
Fig.~\ref{fig:airfly_crosscheck_surface} shows the variation of $Y_0(\lambda_0,p_0,T_0)$ calculated from our data when we assume $Y(\lambda,p,T) \propto \mu$, the measured peak position of the \isotope{Po}{212} alpha decays at different $T$ and $p$. 
The roughly constant corrected values (to within \SI{\pm 10}{\%}) across the full range of temperatures and pressures when the detector was entirely in the gas phase validates the use of the \airfly\ correction factor.

\section{Relative $\alpha/\beta$ quenching} 
\label{appendix:relative_quenching}

As can be seen in Fig.~\ref{fig:spectrum_Q0_bkgsub_ln2}, the energy spectrum of \isotope{Bi}{212} events in liquid nitrogen extends up to $\sim$\SI{200}{\PE}. To understand what fraction of this signal is attributable to Cherenkov emission, we estimated the Cherenkov contribution of the \isotope{Bi}{212} $\beta$-decay using the same \textsc{geant4} simulation described in Sec.~\ref{sec:mc}.
Charged particles traveling faster than the local speed of light in a medium with index of refraction $n>1$ produce $N$ Cherenkov photons per unit length traveled ($x$) and per unit wavelength ($\lambda$) according to the formula:
\begin{equation}
    \frac{d^2N}{dxd\lambda} = 2\pi\alpha Z^2\left(1-\frac{1}{\beta^2n^2(\lambda)}\right)\frac{1}{\lambda^2}
    \label{eq:cherenkov}
\end{equation}
where $\alpha$ is the fine structure constant, $Z$ is the charge of the particle in units of electron charge, and $\beta=v/c$ is the relativistic velocity~\cite{fernow1986introduction}. \textsc{geant4} implements this process for materials if an index of refraction is defined. For the simulations, we used a refractive index of \SI{1.21}{}~\cite{JOHNS1937, Liveing01101893} for \liquidn, where the variation of the refractive index over the sensitive wavelength range of the \PMT\ is small.
\textsc{geant4} samples the emitted energies in the \isotope{Bi}{212} beta decay, and then generates optical photons according to Eq.~\ref{eq:cherenkov} as the beta loses energy until $\beta<1/n$. We record every generated photon but do not optically track them after generation. For \isotope{Bi}{212} beta decays near the endpoint ($Q_{\beta}=\qty[]{2252}{\keV}$), the simulation predicts roughly \SI{1350}{} Cherenkov photons generated between \SIrange{150}{700}{\nano\meter} in \liquidn.

The number of fully-amplified photoelectrons generated from each decay is calculated by Bernoulli sampling, using the detection efficiency, \DE$_{\liquidn}$, which includes the averaged light collection efficiency (\LCE) in the detector, the wavelength dependent \PMT\ \QE, and the \PMT\ \CE, and under-amplified photoelectron fraction, $f$ (see Appx.~\ref{appendix:pmt}).
The \LCE\ of the detector is determined from the position of the \isotope{Po}{212} peak ($\mu$) in cold gas and \RT, as follows:
\begin{align}
\LCE    &= P_{det} / P_{gen} \\
P_{det}       &= \frac{\mu}{\QE(T) \cdot \CE \cdot f} \\
P_{gen}       &= Y(p,T) \cdot E_\alpha \nonumber \\ 
        &= Y_{\gasn}^{\STP} \cdot \Theta(p,T) \cdot E_\alpha 
\end{align}
where $P_{det}$ is the mean number of photons reaching the \PMT, and $P_{gen}$ is the mean number of generated photons. 
$P_{gen}$ is calculated by multiplying the scintillation yield at $(p,T)$, obtained by correcting $Y_{\gasn}^{\STP}$ using $\Theta(p,T)$, by the energy $E_\alpha$ of the alpha particle.
From the results of $\mu$ in Sec.~\ref{sec:mc}, we obtain a volume-averaged \LCE\ of \LCEgas\ at cold, and \LCEgasRT\ at \RT. Note that the \LCE\ is calculated using the gas scintillation signal, which is most intense at \SI{337}{\nano\meter}, but the Cherenkov spectrum is peaked in the UV region. However, since we do not have any other calibration point, we assume that the \LCE\ (taken as \SI{80}{\percent}, the average of the RT and cold gas numbers) is the same for Cherenkov emission in liquid and scintillation in gas.

Given these assumptions, \isotope{Bi}{212} beta decays at the endpoint are estimated to produce a Cherenkov signal of roughly \SIrange{95}{140}{\PE} which is well above the trigger threshold. Assuming that the rest of the measured \isotope{Bi}{212} signal comes from the simultaneous emission of scintillation light, we can use this estimate to calculate a relative scintillation quenching factor between alpha and beta decays. With reference to Fig.~\ref{fig:spectrum_Q0_bkgsub_ln2} which shows a beta-like spectrum with endpoint at roughly \SI{200}{PE}, we estimate that the \SI{2252}{\keV} betas produce roughly \SIrange{60}{105}{\PE} from scintillation. This corresponds to a signal yield of \SIrange{27}{47}{\PE\per\MeV}, compared to \SI{0.6}{\PE\per\MeV} for alphas, implying a quenching factor of more than \SI{40}{}.

\bibliography{bibliography.bib}

\end{document}